\documentclass[format=acmsmall, review=false, screen=true]{acmart}

\usepackage{booktabs} 

\usepackage[ruled]{algorithm2e} 

\SetAlFnt{\small}
\SetAlCapFnt{\small}
\SetAlCapNameFnt{\small}
\SetAlCapHSkip{0pt}
\IncMargin{-\parindent}

\usepackage[font=small,labelfont=bf,tableposition=top]{caption}

\usepackage{pgfplots}
\pgfplotsset{compat=1.13} 
\usetikzlibrary{pgfplots.groupplots}
\usetikzlibrary{shapes.geometric}
\selectcolormodel{cmyk}

\pgfplotsset{
    jitter/.style={
        y filter/.code={\pgfmathparse{\pgfmathresult+rnd*#1}}
    },
    jitter/.default=0.05
}

\newcommand{\nop}[1]{}

\newcommand*{\eg}{{\em e.g.}}

\newcommand*{\ie}{{\em i.e.}}

\setcopyright{acmlicensed}

\received{April 2018} 
\received[revised]{July 2018}
\received[accepted]{September 2018}

\begin{document}
\title[GuessTheKarma]{GuessTheKarma: A Game to Assess Social Rating Systems}

\author{Maria Glenski}
\affiliation{%
  \institution{University of Notre Dame}
  \streetaddress{~}
  \city{Notre Dame}
  \country{United States}}
\email{mglenski@nd.edu}

\author{Greg Stoddard}
\affiliation{%
  \institution{University of Chicago Crime Lab}
  \streetaddress{~}
  \city{New York City}
  \country{United States}}
\email{stoddardg@gmail.com}

\author{Paul Resnick}
\affiliation{%
  \institution{University of Michigan}
  \streetaddress{~}
  \city{Ann Arbor}
  \country{United States}}
\email{presnick@nd.edu}

\author{Tim Weninger}
\affiliation{%
  \institution{University of Notre Dame}
  \streetaddress{~}
  \city{Notre Dame}
  \country{United States}}
\email{tweninge@nd.edu}

\begin{abstract}
Popularity systems, like Twitter retweets, Reddit upvotes, and Pinterest pins  have the potential to guide people toward posts that others liked. That, however, creates a feedback loop that reduces their informativeness: items marked as more popular get more attention, so that additional upvotes and retweets may simply reflect the increased attention and not independent information about the fraction of people that like the items. How much information remains? For example, how confident can we be that more people prefer item A to item B if item A had hundreds of upvotes on Reddit and item B had only a few?
We investigate using an Internet game called GuessTheKarma that collects independent preference judgments (N=20,674) for 400 pairs of images, approximately 50 per pair. Unlike the rating systems that dominate social media services, GuessTheKarma is devoid of social and ranking effects that influence ratings. Overall, Reddit scores were not very good predictors of the true population preferences for items as measured by GuessTheKarma: the image with higher score was preferred by a majority of independent raters only 68\% of the time. However, when one image had a low score and the other was one of the highest scoring in its subreddit, the higher scoring image was preferred nearly 90\% of the time by the majority of independent raters. Similarly, Imgur view counts for the images were poor predictors except when there were orders of magnitude differences between the pairs. We conclude that popularity systems marked by feedback loops may convey a strong signal about population preferences, but only when comparing items that received vastly different popularity scores.

\end{abstract}

\setcopyright{acmlicensed}
\acmJournal{PACMHCI}
\acmYear{2018} \acmVolume{2} \acmNumber{CSCW} \acmArticle{59} \acmMonth{11} \acmPrice{15.00}\acmDOI{10.1145/3274328}

%
%



\maketitle

\renewcommand{\shortauthors}{M. Glenski et al.}

\noindent People increasingly rely on social media as a major source of news, information, and entertainment. Although social media platforms differ in the way that information is curated and delivered, user ratings (\eg, likes, upvotes, pins) play a significant role in determining what is considered to be popular or trending. Tasked with curating an ever-increasing amount of content, providers leverage the collective ratings of the crowd, and measures of collective attention, to identify which content to show users. Unfortunately, peer recommendation can result in irrational herding, where upvotes beget more upvotes and downvotes may cause comments or posts to go unnoticed by the crowd~\cite{lampe2004slash} or where identical content receives widely different attention~\cite{gilbert2013widespread,lakkaraju2013s}.

Social recommendation systems are built on the assumption that collective opinions result in high quality judgments, even better than the judgments of experts~\cite{surowiecki2005wisdom}. However, the so-called wisdom of the crowd tends to work best when ratings are performed independently~\cite{ladha1995information}. When ratings are not performed independently, as in, for example, social media systems, social influence dynamics such as \textit{herding}~\cite{muchnik2013social,glenski2017rating,salganik2006experimental}, \textit{information cascades}~\cite{dow2013anatomy}, and the \textit{ranking bias effect}~\cite{lerman2014leveraging} significantly bias individual judgment~\cite{lorenz2011social,van2014field}. Likewise, the way an item is described has a tremendous impact on its popularity within a community~\cite{lakkaraju2013s,tan2014effect}. Because individuals' perceptions of quality follow the behavior of a group~\cite{janis1982groupthink}, the content made popular by online rating systems may be only weakly correlated with its intrinsic quality~\cite{burghardt2016myopia,bikhchandani1992theory,tan2014effect}. 

Other results on modern social rating systems such as Facebook, Reddit, and Imgur~\cite{stoddard2015popularity} as well as in online education systems~\cite{celis2016sequential} argue that the popularity of a post is a relatively strong reflection of its intrinsic quality. 
Figueiredo et al, in a study somewhat analogous to ours, had mixed results~\cite{Figueiredo:2014:CDI:2611105.2557285}. They asked 6-10 MTurk workers to pick which of two YouTube videos they preferred. YouTube view counts were predictive of reported preferences when there was a clear consensus among the Turkers. However, a large discrepancy in YouTube view counts for a pair of videos was not predictive of whether there would be consensus: in other words, high popularity was not a reliable indicator of consensus preference.
So what do we make of these contradictory reports? Do social influence dynamics fatally distort the relationship between measured popularity and true preferences in social media? 
 
It is easy to tell when something is popular: we simply count the number of views, votes, mentions, etc. On the other hand, it is difficult (perhaps impossible) to determine the objective quality of some content -- perhaps no such thing exists~\cite{arapakis2014feasibility}. 
Our goal is to remove the social influence dynamics as well as the algorithmic and design biases that exist in most social rating systems and solicit independent judgments in the absence of such factors -- \ie, true preferences. We examine the relationship between path-independent judgments of user preference and the respective path-dependent popularity on live social rating systems. Similar outcomes would be evidence that social rating systems promote the content that users collectively prefer. Dissimilar outcomes would be evidence to the contrary.

Methods to solicit ratings from social media users vary significantly. For instance, online purchases and ride-sharing interactions are often measured on a scale from 0 to 5-stars; however, comparing judgments across items and users can be problematic because each person has a different idea of what constitutes a 5-star rating~\cite{masthoff2003modeling}. To relieve this so-called calibration problem, many social surveys ask users to pick the better of a pair of items or the `best' out of a handful of options.

People who care more about an item, positively or negatively, will be more drawn to rating it. This creates a selection bias. At the extreme of selection bias, popularity estimates can be manipulated by actively recruiting participants to rate particular items. Selection bias can be reduced by having the system assign things for people to rate rather than letting people choose for themselves~\cite{SalganikWikiSurveys}.

While assigning items to raters reduces opportunities for deliberate manipulation or inadvertent selection bias, it may reduce motivation for people to participate. It also does nothing to provide incentives for effort in evaluating items or honest reporting of answers. Games have been developed that randomly pair two users and award points if they both pick the same object as the `best'~\cite{hacker2009matchin,von2004labeling}. This motivates participation because of the challenge and collaboration involved and encourages users to take the task seriously because they are only awarded points if they choose accurately. 

Unfortunately, simply rewarding users for matching responses of others creates an incentive to report "focal points" that may not be very informative; if you suspect that others are unlikely to notice a detail about an item, you have an incentive not to base your rating on that detail. Peer prediction techniques respond to this problem by converting people's ratings into predictions about the distribution of ratings that peers will provide and awarding points based on the match of that implied distribution to actual peer ratings rather than based on simply matching the reports~\cite{PeerPrediction} or by rewarding matches on individual ratings but penalizing for blind agreement across items~\cite{Dasgupta:2013:CJE:2488388.2488417}. 

Another approach to encourage honest reporting is the Bayesian Truth Serum (BTS): each user is asked to report their own rating and to predict the ratings of others concurrently; a scoring rule awards points in a way that rewards honest reporting of both values~\cite{Prelec462}.
These surveys have been shown to return high-quality ratings~\cite{shaw2011designing}. A study comparing product ratings collected traditionally versus using a BTS approach found that the BTS-ratings converged to the traditional ratings but did so faster and with far less variance than the traditional survey~\cite{gilbert2014if}.

In the present work we introduce and present the results of an Internet game called {\em GuessTheKarma} that asked users to select, from a pair of images, which they personally prefer and to guess which image had received more votes when previously posted in an image-based subreddit on Reddit. The personal preference reports provide a proxy for the groundtruth popularity, the aggregate preference of all potential viewers. To prevent selection effects, the system chose pairs to present rather than allowing users to choose. We did not use a Peer Prediction or BTS scoring system to incentivize honest reporting because we lacked the mechanism to solicit peer ratings. The primary role of asking users to predict which image scored best on Reddit was to make participation fun; once they were participating, players had no reason not to honestly report which images they preferred. 

We use the groundtruth determined by the GuessTheKarma player's reported preferences to assess the informativeness of other available popularity measures based on self-selected raters and user attention. One such popularity measure is the Reddit vote-score, essentially the number of upvotes minus the number of downvotes an item received from Reddit users. The score of a post on Reddit can be influenced by many factors other than the population's true preference for the item. A post's submission time, the number of early votes, its relative position, the subreddit to which it is submitted, and other factors are all important in determining a post's fate. GuessTheKarma eliminates these social, visual, and algorithmic biases and instead asks human judges to focus solely on comparing the content of the paired images.

A second, more passive, measure of popularity is viewership. Many images posted to Reddit are hosted on Imgur, which reports viewership statistics of each hosted image. We can employ Imgur viewership statistics as another popularity predictor. Because of its algorithmic and presentation design choices, Imgur is likely to have different types of socio-technical biases, which can provide a different perspective on the same content. 

The goal of this work is to assess the ability of social media popularity metrics to highlight informative or interesting content. This assessment is performed by using the results of the GuessTheKarma game as groundtruth for content preference, and treating the social media outcomes, \ie, Imgur views and Reddit scores, as predictors which come from a live, influence-rich environment. Prediction accuracy can then be measured against the GuessTheKarma groundtruth and used as evidence of the ability of social media platforms to identify preferred content.

Our results show that Reddit scores and Imgur views are surprisingly poor predictors of user preference. Each image pair was evaluated by about 50 GuessTheKarma players. The image with a higher Reddit score or more Imgur views was preferred by the majority of the players around two-thirds of the time. 
Conversely, GuessTheKarma players were able to predict which image was preferred only about 60\% of the time, and self-reported Reddit-powerusers were no better than non-users.  Additional analyses investigated the effect of player agreement and subreddit size. From this analysis we were able to distill circumstances that resulted in good predictive performance.

\smallskip
\noindent
In summary, our contributions are two-fold:
\begin{enumerate}
    \item 
    GuessTheKarma, a game-like survey tool that can be used to explore the performance of social rating systems, and
    \item 
    Analysis of whether Reddit vote-scores and Imgur view-counts are predictors of the true majority opinion, when that opinion is  gathered without social influence.
\end{enumerate}

\section{Methodology}

\subsection{GuessTheKarma Design} 
  
\begin{figure}[t]
\small
\centering
\begin{tabular}{c}
\includegraphics[width=\linewidth]{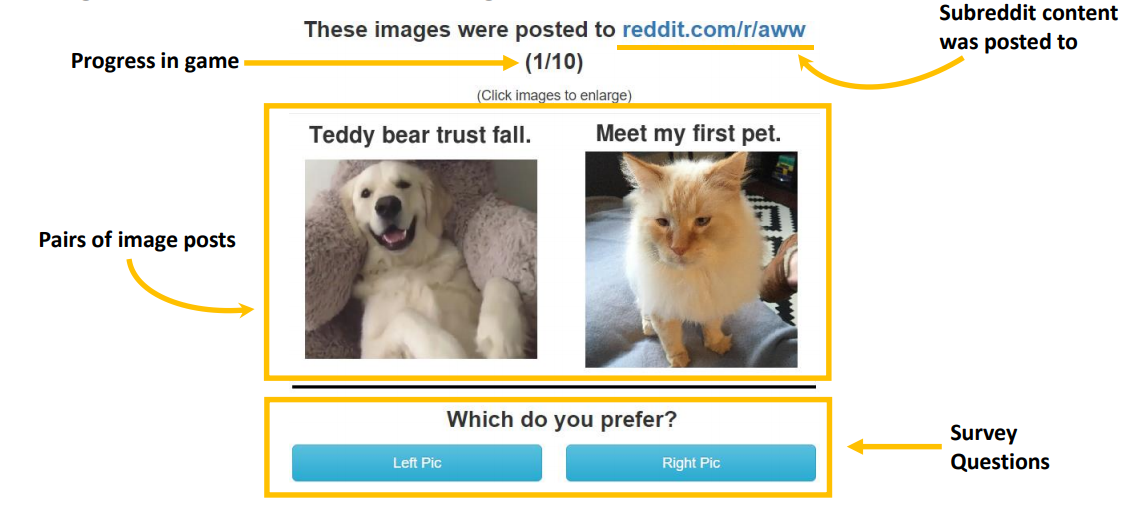}\\

\hspace{-.25in}
\rule[0.5ex]{2cm}{0.5pt}~~Click~~\rule[0.5ex]{2cm}{0.5pt}\\

\hspace{-.25in}
\includegraphics[width=.46\linewidth]{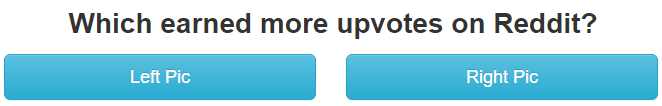}\\

\hspace{-.25in}
\rule[0.5ex]{2cm}{0.5pt}~~Click~~\rule[0.5ex]{2cm}{0.5pt}\\


\hspace{-.25in}
\includegraphics[width=.46\linewidth]{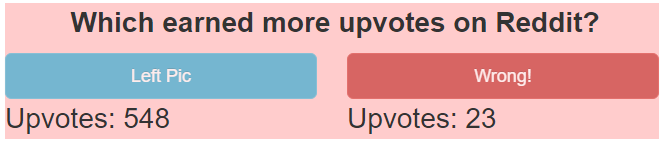}\\
\end{tabular}
\vspace{-0.5\baselineskip}
\caption{Example of a single round of GuessTheKarma with main design aspects highlighted wherein the change in survey question after a selection is made is illustrated after ``--- Click ---''.} 
\vspace{-1\baselineskip}
\label{fig:gtk_design}
\end{figure}

GuessTheKarma is a single-player online game that asks players which of a pair of image-posts they prefer, \ie, find more interesting or informative. The game is straightforward: immediately upon entering the Web page, the player is randomly assigned one of eight subreddits (r/aww, r/pics, r/funny, r/OldSchoolCool, /r/photocritique, /r/CrappyDesign, /r/itookapicture, or /r/EarthPorn) from which images will be selected. For example, if the subreddit /r/funny is selected, then all images in the game would be drawn from image-posts submitted to /r/funny. Then, without further instruction, each player is presented with two image-posts and the survey question(s). Figure~\ref{fig:gtk_design} illustrates a single round of the game, wherein the survey-text changes when a player makes a choice, denoted by -- Click --. These eight subreddits were chosen because they primarily host images and because their subscriber count, \ie, viewership, varies from very large to very small. Previous work has found that post titles have a dramatic effect on the final popularity of Reddit images~\cite{lakkaraju2013s}. The post title is integral to the content of the post, so we include it above the image. The player may choose a different subreddit manually at anytime but, if that occurs, the game will restart using the new subreddit. 

Upon making their selection(s), the final scores of both posts are displayed (\ie, the number of upvotes minus the number of downvotes received), the area around the decision-buttons turns green, and the button-text changes to ``Correct!'' if the player guesses correctly. If the player guesses incorrectly, the area around the decision-buttons turns red and the button-text changes to ``Wrong!''. Figure~\ref{fig:gtk_design} illustrates the result of an incorrect choice.

After displaying the score totals for 3 seconds, the game automatically progresses to the next pair of images. The counter at the top of the page increments with every selection until 10 image-pairs have been judged by the player. At any time the user may manually select a different subreddit, but this restarts the game. Partially completed games are not recorded. After the final image-pair, but before we showed a player their accuracy, we asked each player to answer a short usage questionnaire shown in Tab.~\ref{tab:questionnaire_gtk2}

There is an important distinction between asking ``what do you prefer?'' and asking ``what do you think others prefer?'' Hacker and Von Ahn compare this distinction to the problem of election polling~\cite{hacker2009matchin}, where a similar distinction exists in the questions: ``who will you vote for?'' versus ``who do you think will win?'' respectively. It is unclear which question better elicits information about an image post but a study on political polling showed that asking voters who they think will win, rather than who they will vote for, is a better predictor of the actual election winner~\cite{rothschild2011forecasting}. Nevertheless, the goal of this work is to assess how well social media systems aggregate ratings from social media users, so we use the preference question to formulate a groundtruth of user preferences. In our study, the prediction question served mainly to motivate participation: it was challenging and fun to try to predict which image got more votes on Reddit.

Despite the wording of the prediction-judgment survey question, which was chosen based on feedback from beta-testers of the game, we judge correctness based on the \textit{score} of the post not the number of upvotes. Unfortunately, it is impossible to know the actual number of upvotes on any Reddit post through Reddit's API. We chose to use the word ``upvote'' instead of ``score'' because 1) asking for the number of upvotes is more clear -- many game players do not understand the details of how the score of a post is calculated, and 2) there is little distinction anyways -- the post with the most upvotes will normally also have the higher score.

We collected image-posts for each subreddit from the Reddit BigQuery dataset\footnote{Available at \url{https://bigquery.cloud.google.com/dataset/fh-bigquery:reddit_posts}}, which contained  posts submitted between January 1, 2008 and August 31, 2015. Rather than randomly pairing images, we separated image posts into very high-scoring (VH), high-scoring (H), medium-scoring (M), and low-scoring (L) bins, which correspond to posts with a final score $>$ 95\%, between 75-95\%, 50-75\%, and $<$ 50\% of all posts within their respective subreddits. Then, when we created the sets of image-pairs to display to players, we randomly sampled image-pairs from 6 bin-pair-permutations to create games using only VH--VH, H--H, VH--H, H--M, H--L, and VH--L pairings. Images were placed on the left or right randomly. Reposts of the same image and post-title are possible resulting in different scores for the same data. In these cases, we select the highest scoring post.

Of the $4^2$ possible combinations, we purposely limited the image pairings to those that contained at least one high- or very high-scoring image. In beta-tests, players complained that L--L, M--L, etc. pairings were too difficult and unfair. These complaints were reasonable; an image-pair with a score of 5 versus 4 is essentially a toss up. Furthermore, the goal of the game is to assess whether those images which became popular on social media match a path-independent groundtruth based on GuessTheKarma's survey of user preference.
We used 400 total combinations of 325 image-posts and added safeguards to the survey so that a player does not see the same image-pair twice. Limiting the number of image-posts resulted in each pair receiving judgments from multiple players. Multiple judgments on the same image-pair permit an assessment using a crowd-sourced majority-opinion -- our groundtruth.

The GuessTheKarma methodology was approved by the University of Notre Dame's Institutional Review Board (\#17-06-3941).

\subsection{Data Collection}
 
On February 29, 2017, we opened the Web site and recruited players. We recruited game players via posts to Reddit, Digg, Twitter, and various other social networks. It is difficult to determine precisely, but we estimate that the majority of the players arrived via Reddit. By March 7, 2017, 2,660 people had played the game and provided 20,674 preference judgments. We only recorded votes after the player made judgments for all 10 pairs.  Table~\ref{tab:bigdata2} shows the distribution of judgments and distinct images used across subreddits. Of the 2,660 total players, 2,083 (78.3\%) completed the usage questionnaire at the end of the survey. Their responses are displayed in Table~\ref{tab:questionnaire_gtk2}. 

\begin{table}[t]
\small
\centering
\caption{Guess The Karma Dataset.}
\vspace{-0.75\baselineskip} 
\begin{tabular}{lrrr}
\hline
      Subreddit &  \# Judgements &  \# Image Pairs &  \# Images \\
\hline
         /r/funny &          3,090 &             50 &           42 \\
 /r/OldSchoolCool &          3,460 &             50 &           42 \\
           /r/aww &          3,334 &             50 &           41 \\
          /r/pics &          3,280 &             50 &           42 \\
 /r/photocritique &          1,130 &             50 &           42 \\
  /r/CrappyDesign &          1,726 &             50 &           35 \\
 /r/itookapicture &          3,170 &             50 &           42 \\
     /r/EarthPorn &          1,484 &             50 &           39 \\ 
\hline
\textbf{Total}    &\textbf{20,674} &   \textbf{400} & \textbf{325} \\
\hline 
\end{tabular}
\label{tab:bigdata2}
\end{table}

\begin{table}[t]
\small
\centering
\caption{Summary of the responses to the Reddit usage questionnaire.}
\vspace{-0.75\baselineskip}
\begin{tabular}{l|lll}
\hline
Question & \multicolumn{3}{c}{Distribution of Responses} \\ \hline
Describe your level of Reddit use & Heavy & 	Casual & 	Don't use Reddit \\
&1294 (59.9\%)& 	724 (33.5\%)& 	144 (6.7\%)  \\	
How long have you used Reddit? & Over a year & 	 0-12 months  & 	Don't use Reddit \\ 	
&1815 (84.0\%)& 	202 (9.3\%)& 	145 (6.7\%)  \\	
Do you pay attention to r/$x$?& Yes & 	No & 	Don't use Reddit \\	
&606 (28.0\%)& 	1227 (56.8\%)& 	329 (15.2\%)  	\\
Do you vote on posts?& Yes & 	No & 	Don't use Reddit \\ 	
&1371 (63.4\%)& 	643 (29.7\%)& 	148 (6.8\%) \\	
Do you vote on posts in reddit.com/r/new?& Yes & 	No & 	Don't use Reddit \\	
&279 (12.9\%)& 	1674 (77.4\%)& 	209 (9.7\%)\\	
\end{tabular}
\vspace{-0.75\baselineskip}
\label{tab:questionnaire_gtk2}
\end{table}

Sampling bias is a fundamental limitation of many Web-based studies; GuessTheKarma is no exception. We attempted to limit this issue by directly recruiting participants from Reddit (rather than Mechanical Turk) but our particular sample of users may not be completely representative of all social media users or all persons. For example, many of our participants came from a post submitted to the /r/webGames subreddit, but it is unclear how the demographics of the /r/webGames subreddit reflect the general demographics on Reddit or the demographics of the users of the image-subreddits used to populate the games.

\subsection{Analysis}

In this section, we outline how we compare accuracy of real-world, socially-aggregated outcomes (Reddit vote-scores and Imgur view-counts) against GuessTheKarma preferences. The GuessTheKarma outcome of an image-pair is measured by the majority preference of GuessTheKarma players. As discussed earlier, we view the majority preference as a kind of unbiased groundtruth to which other outcomes are compared. Before we examine accuracy against the groundtruth measure, we first examine the agreement among the GuessTheKarma players. 

We use Fleiss' Kappa ($\kappa$) to measure the agreement of judgments for an image-pair where 1 is complete agreement and 0 is agreement indistinguishable from random. Negative $\kappa$ scores, therefore, can be interpreted as being worse than random agreement. 

Next, we examine the overall accuracy of our two predictors. First, we want to know: how accurately do social aggregators like Reddit and Imgur predict the majority opinions of ``the crowd''? These predictors rely on judgments collected from an environment with social influence effects, as well as rank, design, and other algorithmic-biases; poor prediction accuracy may therefore be attributed to the presence of these biases. We compare the accuracy and 95\% confidence intervals for each predictor and perform tests of statistical significance. We also look for correlations between the predictor accuracy and groundtruth agreement, the choice of subreddit, the Reddit-score-percentile differences within image-pairs, and player expertise.

\medskip
\noindent \textbf{Agreement Effect}
Accuracy may be affected by the level of agreement for the pair of image-posts. We hypothesize that image-pairs with complete agreement ($\kappa=1$) would be \textit{easier} to predict than those with low ($\kappa\approx0$) agreement. To answer this question, we plot the accuracy of each image-pair as a function of their agreement. We then perform statistical tests to analyze what correlation, if any, exists.

\medskip
\noindent \textbf{Subreddit Effect}
The choice of subreddit may also affect the accuracy of popularity predictions. 
Across the eight subreddits, the size of the community, \ie, the number of subscribers to the subreddit, varies widely; for example, as of February 23, 2017, /r/pics had 18.4 million subscribers while /r/photocritique had only 36 thousand subscribers. The /r/photocritique, /r/CrappyDesign, /r/itookapicture, and /r/EarthPorn subreddits also have a much narrower focus for the images submitted than r/aww, r/pics, r/funny, or r/OldSchoolCool. We compare the accuracy of platform predictors for each subreddit.

\medskip
\noindent \textbf{Image-pair Score Balance}
Recall that images were labeled as being VH, H, M, and L scoring based on their subreddit-conditioned score percentile. Images were purposefully paired according to their labels. We call image-pairs comprised of similar scoring posts balanced, and image-pairs with dissimilar scores unbalanced.  We expect that it is easier to predict the correct image from an unbalanced image-pairing (i.e., where one image received far more attention than the other), than from a balanced image-pair (where the two images receive about the same amount of attention). We test this hypothesis by plotting platform accuracy with respect to the score difference of the image-pair.

\medskip
\noindent
Data, the GuessTheKarma source code, and the complete statistical analysis scripts are available online at \url{https://github.com/nddsg/GTK-paper}.

\section{Results} 

First, we examine the overall accuracy of our platform predictors against the majority preference of the GuessTheKarma players. Accuracy is measured by comparing the predictor choice, \ie, which post received the higher Reddit-score or more Imgur-views, against the GuessTheKarma majority preference. 

The score of Reddit posts and the number of views on Imgur was highly correlated ($R^2=0.80$, $p<0.001$); of the 400 image-pairs in our data set, the image with the higher Imgur-popularity matched the post with the higher Reddit score 86.3\% of the time. Thus, the informativeness of both measures was similar. 

We report 95\% confidence intervals around the accuracy estimates. Reddit vote-scores and Imgur view-counts had accuracies of 68.0\% $\pm$ 4.6\% and 64.7\% $\pm$ 4.7\%, respectively.

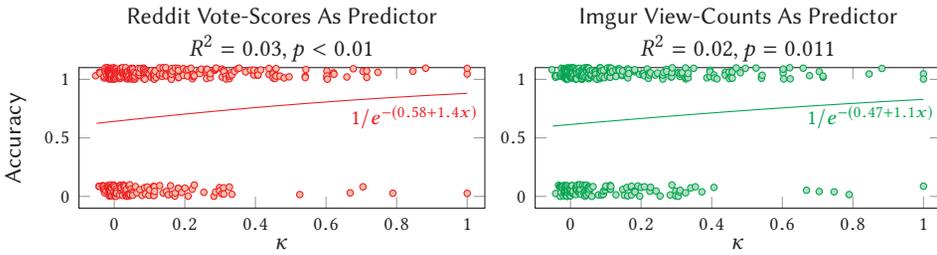
\begin{figure}[t]
\centering 
\begin{tikzpicture}
\sffamily
\begin{axis}[  
    title = {\small Reddit Vote-Scores As Predictor \\\small $R^2=0.03$, $p<0.01$}, 
    title style={align=center,yshift=-.1in}, 
    legend style={font=\scriptsize},
    width = 2.75in, height = 1.35in, 
    ylabel near ticks,
    xlabel near ticks,
    ylabel = {\small Accuracy}, 
    ymax=1.1, ymin=-.1,
    every tick label/.append style={font=\scriptsize},
    xmin=-0.1,xmax=1.05, 
    domain=-0.1:2, 
    xlabel={\small $\kappa$},
    xlabel style = {yshift=0.05in},
    ] 
     
    \addplot[no marks, solid, red, domain=-0.05:1]{(1/(1+exp(-(0.5781+1.4167*x))))};
    \node[red] at (axis cs: .85,.68) {\footnotesize{$1/e^{-(0.58+1.4x)}$}};
      \addplot[only marks,mark options={scale=.6}, red, fill=red!30!white, jitter=0.1] coordinates {
(-0.052631578947368474,1)
(-0.04347826086956519,0)
(-0.04210526315789476,1)
(-0.04210526315789476,0)
(-0.040000000000000036,1)
(-0.03384615384615386,0)
(-0.030303030303030276,0)
(-0.0292397660818714,1)
(-0.0292397660818714,0)
(-0.0292397660818714,0)
(-0.0292397660818714,0)
(-0.02857142857142858,1)
(-0.026666666666666616,0)
(-0.02564102564102566,1)
(-0.02564102564102566,0)
(-0.02564102564102566,1)
(-0.02564102564102566,0)
(-0.02564102564102566,1)
(-0.023076923076923106,0)
(-0.022727272727272707,1)
(-0.020242914979757054,0)
(-0.020242914979757054,1)
(-0.020242914979757054,1)
(-0.019607843137254943,1)
(-0.018181818181818188,0)
(-0.018181818181818188,1)
(-0.018181818181818188,1)
(-0.016470588235294126,0)
(-0.016129032258064502,1)
(-0.016042780748663055,0)
(-0.015819209039548032,1)
(-0.01548821548821544,1)
(-0.01548821548821544,0)
(-0.01548821548821544,0)
(-0.01538461538461533,0)
(-0.01538461538461533,1)
(-0.01538461538461533,1)
(-0.01538461538461533,1)
(-0.015335801163405605,0)
(-0.015335801163405605,0)
(-0.015335801163405605,1)
(-0.014207650273224015,1)
(-0.014207650273224015,1)
(-0.014207650273224015,1)
(-0.014047410008779626,0)
(-0.014047410008779626,0)
(-0.013664596273291973,1)
(-0.013664596273291973,1)
(-0.013664596273291973,1)
(-0.013461538461538414,1)
(-0.012429378531073398,1)
(-0.012429378531073398,0)
(-0.012162876784770016,1)
(-0.011891891891891881,1)
(-0.011413520632133412,0)
(-0.011413520632133412,0)
(-0.011180124223602483,0)
(-0.011180124223602483,1)
(-0.010954616588419452,1)
(-0.010954616588419452,0)
(-0.010736764161421664,1)
(-0.010526315789473717,1)
(-0.010526315789473717,1)
(-0.010526315789473717,1)
(-0.010196078431372602,1)
(-0.009935710111046214,0)
(-0.009935710111046214,0)
(-0.009935710111046214,0)
(-0.009935710111046214,0)
(-0.00983606557377048,1)
(-0.00983606557377048,1)
(-0.00983606557377048,0)
(-0.009753298909925379,0)
(-0.009615384615384581,1)
(-0.009615384615384581,1)
(-0.009446693657219951,1)
(-0.009446693657219951,0)
(-0.0091324200913242,0)
(-0.009009009009009028,1)
(-0.008403361344537785,1)
(-0.008403361344537785,1)
(-0.007575757575757569,1)
(-0.007039337474120111,1)
(-0.0070237050043898686,0)
(-0.006993006993006978,1)
(-0.00694444444444442,1)
(-0.00687466948704385,0)
(-0.00687466948704385,1)
(-0.00687466948704385,1)
(-0.00687466948704385,1)
(-0.006779661016949157,1)
(-0.006779661016949157,0)
(-0.004566210045662156,1)
(-0.003278688524590123,0)
(-0.002020202020202033,0)
(-0.002020202020202033,0)
(-0.002020202020202033,1)
(-0.002020202020202033,1)
(-0.001851166234727919,1)
(-0.001851166234727919,1)
(-0.001851166234727919,1)
(-0.0015649452269170805,1)
(-0.0015649452269170805,0)
(-0.001242236024844745,1)
(-0.0007843137254901489,0)
(-0.0004662004662004948,1)
(0.0,0)
(0.0005288207297726721,1)
(0.0005288207297726721,1)
(0.0006835269993163973,1)
(0.001129943502824915,1)
(0.001129943502824915,1)
(0.0017825311942958333,0)
(0.0017825311942958333,0)
(0.0022948938611588865,1)
(0.0022948938611588865,1)
(0.0038461538461538325,1)
(0.0039525691699604515,1)
(0.0039525691699604515,1)
(0.0039525691699604515,1)
(0.0040322580645162365,0)
(0.0048130322102923895,0)
(0.0048130322102923895,1)
(0.005464480874316946,1)
(0.006211180124223503,0)
(0.006211180124223503,1)
(0.006428988895382792,0)
(0.006428988895382792,1)
(0.006428988895382792,0)
(0.006747638326585648,1)
(0.006815968841285214,0)
(0.0070237050043897575,0)
(0.008288288288288204,1)
(0.010047593865679438,1)
(0.0117647058823529,1)
(0.0117647058823529,1)
(0.014084507042253502,1)
(0.015151515151515138,0)
(0.015151515151515138,1)
(0.015151515151515138,1)
(0.015320910973084967,1)
(0.015320910973084967,1)
(0.01538461538461533,1)
(0.016064257028112428,0)
(0.01754385964912286,1)
(0.01754385964912286,1)
(0.018118059614260718,1)
(0.018118059614260718,0)
(0.018118059614260718,0)
(0.018181818181818077,1)
(0.0182648401826484,0)
(0.01984126984126977,0)
(0.022222222222222143,1)
(0.023693605972086917,1)
(0.023728813559322104,1)
(0.023728813559322104,1)
(0.023728813559322104,0)
(0.02499999999999991,1)
(0.025290498974709585,1)
(0.025290498974709585,0)
(0.027027027027026973,0)
(0.027450980392156765,1)
(0.027450980392156765,0)
(0.028094820017559252,0)
(0.028340080971659853,1)
(0.028919330289193246,1)
(0.030303030303030276,0)
(0.03214494447691418,0)
(0.032258064516129004,1)
(0.032258064516129004,1)
(0.032258064516129004,0)
(0.032738095238095344,1)
(0.03407984420642651,1)
(0.03442340791738374,1)
(0.03442340791738374,0)
(0.03442340791738374,0)
(0.03543098889476459,1)
(0.03599374021909241,1)
(0.03599374021909241,0)
(0.03838383838383841,1)
(0.03850931677018643,1)
(0.03850931677018643,1)
(0.03855855855855861,0)
(0.040000000000000036,1)
(0.041095890410958846,0)
(0.0412642669007901,0)
(0.042105263157894646,1)
(0.042105263157894646,0)
(0.042105263157894646,0)
(0.04428904428904423,1)
(0.04428904428904423,0)
(0.04480874316939887,0)
(0.045324153757888785,1)
(0.045324153757888785,0)
(0.04627450980392167,1)
(0.04627450980392167,1)
(0.04761904761904767,1)
(0.04761904761904767,0)
(0.04850964348334297,1)
(0.04850964348334297,1)
(0.048530416951469535,1)
(0.05129561078794298,0)
(0.05129561078794298,1)
(0.05138339920948609,1)
(0.05138339920948609,1)
(0.05138339920948609,1)
(0.05153153153153145,1)
(0.05153153153153145,1)
(0.05258799171842643,1)
(0.05258799171842643,1)
(0.05258799171842643,1)
(0.05258799171842643,0)
(0.05258799171842643,0)
(0.05384615384615388,1)
(0.056189640035118504,1)
(0.05723905723905731,1)
(0.058461538461538565,1)
(0.05882352941176472,1)
(0.05882352941176472,1)
(0.06034801925212885,0)
(0.062200956937799035,1)
(0.06416275430359941,0)
(0.06832298136645965,1)
(0.06832298136645965,1)
(0.07115384615384612,1)
(0.07115384615384612,1)
(0.0730282375851996,0)
(0.07457627118644061,1)
(0.07878787878787885,0)
(0.08333333333333326,1)
(0.08333333333333326,1)
(0.08333333333333326,1)
(0.08333333333333326,1)
(0.08333333333333326,1)
(0.08571428571428563,1)
(0.08571428571428563,0)
(0.08675799086757996,1)
(0.08771929824561409,1)
(0.08860759493670889,0)
(0.08937070333157049,1)
(0.09333333333333327,1)
(0.09333333333333327,1)
(0.09333333333333327,0)
(0.09604519774011289,1)
(0.09859154929577474,0)
(0.0990990990990992,1)
(0.10040160642570273,0)
(0.10476190476190483,1)
(0.10502283105022836,1)
(0.11158117398202005,1)
(0.11158117398202005,1)
(0.11163062536528345,0)
(0.11163062536528345,1)
(0.11163062536528345,0)
(0.11260504201680677,0)
(0.11260504201680677,1)
(0.11260504201680677,1)
(0.11384615384615393,0)
(0.11384615384615393,1)
(0.11384615384615393,1)
(0.11462450592885376,0)
(0.11578947368421044,1)
(0.11888111888111896,1)
(0.12786885245901636,1)
(0.1287878787878789,0)
(0.1287878787878789,1)
(0.1287878787878789,1)
(0.13590692755156009,1)
(0.13734658094681462,1)
(0.1384615384615384,1)
(0.1384615384615384,0)
(0.1457627118644067,1)
(0.1461187214611872,1)
(0.14782608695652177,1)
(0.14782608695652177,0)
(0.15294117647058814,0)
(0.1595959595959595,0)
(0.15961538461538471,1)
(0.15999999999999992,1)
(0.1619718309859155,1)
(0.16234796404019036,1)
(0.16234796404019036,1)
(0.16532258064516125,1)
(0.16532258064516125,0)
(0.17184265010351973,0)
(0.1729323308270676,0)
(0.1729323308270676,1)
(0.18128654970760238,0)
(0.18128654970760238,1)
(0.18128654970760238,1)
(0.18153846153846165,1)
(0.18181818181818188,0)
(0.18205128205128207,1)
(0.18205128205128207,1)
(0.1827027027027026,1)
(0.1909042834479111,0)
(0.19579193454120403,1)
(0.19617224880382778,1)
(0.20029618659755655,0)
(0.2045197740112994,1)
(0.20734126984126977,1)
(0.20934761441090566,0)
(0.20934761441090566,1)
(0.21727395411605932,1)
(0.22077922077922074,0)
(0.22157588577472231,1)
(0.22331002331002336,1)
(0.22331002331002336,1)
(0.2248447204968944,1)
(0.226936026936027,0)
(0.226936026936027,1)
(0.24,1)
(0.24242424242424243,1)
(0.24657534246575352,1)
(0.25383022774327113,0)
(0.26054054054054054,1)
(0.2615384615384615,1)
(0.2615384615384615,1)
(0.2615384615384615,1)
(0.2635885447106956,1)
(0.27231638418079096,1)
(0.27231638418079096,1)
(0.28248796741947424,0)
(0.2846153846153847,1)
(0.28671328671328666,1)
(0.28724440116845185,1)
(0.28724440116845185,0)
(0.28853754940711474,1)
(0.29824561403508776,0)
(0.29824561403508776,0)
(0.305050505050505,1)
(0.305050505050505,0)
(0.3055555555555556,1)
(0.3096045197740114,0)
(0.30980392156862746,1)
(0.31060606060606055,0)
(0.31060606060606055,1)
(0.31580655631288534,1)
(0.31677018633540377,1)
(0.3262032085561497,1)
(0.3262032085561497,0)
(0.3262823902696985,1)
(0.3262823902696985,1)
(0.33063427800269896,1)
(0.33333333333333326,1)
(0.33333333333333326,0)
(0.3407364114552893,1)
(0.3491525423728814,1)
(0.3538461538461539,1)
(0.37225636523266026,1)
(0.38636363636363646,1)
(0.3954116059379218,1)
(0.3954116059379218,1)
(0.3954116059379218,1)
(0.40639269406392686,1)
(0.4066631411951349,1)
(0.4076923076923078,1)
(0.41512605042016815,1)
(0.4385964912280702,1)
(0.4385964912280702,1)
(0.4490929285449834,1)
(0.4500264410364887,1)
(0.4500264410364887,1)
(0.4556451612903225,1)
(0.4584615384615385,1)
(0.4584615384615385,1)
(0.4584615384615385,1)
(0.4696969696969697,1)
(0.4813559322033898,1)
(0.49550502379693273,1)
(0.5256916996047432,0)
(0.5411605937921729,1)
(0.5411605937921729,1)
(0.5430988894764675,1)
(0.6023391812865497,1)
(0.6023391812865497,1)
(0.6023391812865497,1)
(0.6040404040404039,1)
(0.6210526315789473,1)
(0.622693096377307,1)
(0.6590909090909092,1)
(0.6684491978609626,0)
(0.6773109243697479,1)
(0.7046153846153846,0)
(0.7046153846153846,1)
(0.7153846153846153,1)
(0.7153846153846153,1)
(0.7468926553672317,1)
(0.7894736842105263,0)
(0.8461538461538463,1)
(0.8823529411764706,1)
(1.0,0)
(1.0,1)
(1.0,1)
      };

\end{axis}    
\end{tikzpicture} 
\begin{tikzpicture}
\sffamily
\begin{axis}[  
    title = {\small Imgur View-Counts As Predictor\\\small $R^2=0.02$, $p=0.011$ }, 
    title style={align=center,yshift=-.1in}, 
    legend style={font=\scriptsize},
    width = 2.75in, height = 1.35in, 
    ylabel near ticks,
    xlabel near ticks,
    ymax=1.1, ymin=-.1,
    every tick label/.append style={font=\scriptsize},
    xmin=-0.1,xmax=1.05, 
    domain=-0.1:2,  
    xlabel={\small \textcolor{black}{$\kappa$}},
    xlabel style = {yshift=0.05in},
    ] 
    
    \addplot[no marks, solid, green, domain=-0.05:1]{(1/(1+exp(-(0.4646+1.1098*x))))};
    \node[green] at (axis cs: .85,.68) {\footnotesize{$1/e^{-(0.47+1.1x)}$}};
     
    \addplot[only marks,mark options={scale=.6}, green, fill=green!30!white, jitter=0.1] coordinates { 
    (-0.052631578947368474,1)
(-0.04347826086956519,1)
(-0.04210526315789476,1)
(-0.04210526315789476,0)
(-0.040000000000000036,1)
(-0.03384615384615386,0)
(-0.030303030303030276,0)
(-0.0292397660818714,1)
(-0.0292397660818714,0)
(-0.0292397660818714,0)
(-0.0292397660818714,0)
(-0.02857142857142858,1)
(-0.026666666666666616,0)
(-0.02564102564102566,0)
(-0.02564102564102566,1)
(-0.02564102564102566,1)
(-0.02564102564102566,0)
(-0.02564102564102566,1)
(-0.023076923076923106,0)
(-0.022727272727272707,0)
(-0.020242914979757054,0)
(-0.020242914979757054,0)
(-0.020242914979757054,1)
(-0.019607843137254943,1)
(-0.018181818181818188,1)
(-0.018181818181818188,0)
(-0.018181818181818188,1)
(-0.016470588235294126,1)
(-0.016129032258064502,1)
(-0.016042780748663055,0)
(-0.015819209039548032,1)
(-0.01548821548821544,1)
(-0.01548821548821544,0)
(-0.01548821548821544,0)
(-0.01538461538461533,1)
(-0.01538461538461533,1)
(-0.01538461538461533,1)
(-0.01538461538461533,0)
(-0.015335801163405605,0)
(-0.015335801163405605,0)
(-0.015335801163405605,1)
(-0.014207650273224015,1)
(-0.014207650273224015,0)
(-0.014207650273224015,0)
(-0.014047410008779626,0)
(-0.014047410008779626,0)
(-0.013664596273291973,1)
(-0.013664596273291973,1)
(-0.013664596273291973,1)
(-0.013461538461538414,1)
(-0.012429378531073398,1)
(-0.012429378531073398,0)
(-0.012162876784770016,1)
(-0.011891891891891881,1)
(-0.011413520632133412,0)
(-0.011413520632133412,0)
(-0.011180124223602483,1)
(-0.011180124223602483,1)
(-0.010954616588419452,1)
(-0.010954616588419452,0)
(-0.010736764161421664,1)
(-0.010526315789473717,1)
(-0.010526315789473717,1)
(-0.010526315789473717,1)
(-0.010196078431372602,0)
(-0.009935710111046214,0)
(-0.009935710111046214,1)
(-0.009935710111046214,1)
(-0.009935710111046214,0)
(-0.00983606557377048,0)
(-0.00983606557377048,1)
(-0.00983606557377048,1)
(-0.009753298909925379,0)
(-0.009615384615384581,1)
(-0.009615384615384581,0)
(-0.009446693657219951,1)
(-0.009446693657219951,0)
(-0.0091324200913242,0)
(-0.009009009009009028,1)
(-0.008403361344537785,1)
(-0.008403361344537785,1)
(-0.007575757575757569,1)
(-0.007039337474120111,1)
(-0.0070237050043898686,1)
(-0.006993006993006978,1)
(-0.00694444444444442,1)
(-0.00687466948704385,0)
(-0.00687466948704385,1)
(-0.00687466948704385,0)
(-0.00687466948704385,1)
(-0.006779661016949157,1)
(-0.006779661016949157,0)
(-0.004566210045662156,1)
(-0.003278688524590123,0)
(-0.002020202020202033,1)
(-0.002020202020202033,1)
(-0.002020202020202033,0)
(-0.002020202020202033,0)
(-0.001851166234727919,1)
(-0.001851166234727919,1)
(-0.001851166234727919,1)
(-0.0015649452269170805,0)
(-0.0015649452269170805,1)
(-0.001242236024844745,1)
(-0.0007843137254901489,1)
(-0.0004662004662004948,1)
(0.0,0)
(0.0005288207297726721,1)
(0.0005288207297726721,1)
(0.0006835269993163973,1)
(0.001129943502824915,1)
(0.001129943502824915,1)
(0.0017825311942958333,0)
(0.0017825311942958333,0)
(0.0022948938611588865,0)
(0.0022948938611588865,1)
(0.0038461538461538325,1)
(0.0039525691699604515,1)
(0.0039525691699604515,1)
(0.0039525691699604515,1)
(0.0040322580645162365,0)
(0.0048130322102923895,0)
(0.0048130322102923895,1)
(0.005464480874316946,1)
(0.006211180124223503,0)
(0.006211180124223503,1)
(0.006428988895382792,0)
(0.006428988895382792,1)
(0.006428988895382792,1)
(0.006747638326585648,1)
(0.006815968841285214,1)
(0.0070237050043897575,0)
(0.008288288288288204,1)
(0.010047593865679438,1)
(0.0117647058823529,1)
(0.0117647058823529,1)
(0.014084507042253502,1)
(0.015151515151515138,0)
(0.015151515151515138,0)
(0.015151515151515138,1)
(0.015320910973084967,1)
(0.015320910973084967,1)
(0.01538461538461533,1)
(0.016064257028112428,0)
(0.01754385964912286,1)
(0.01754385964912286,0)
(0.018118059614260718,0)
(0.018118059614260718,1)
(0.018118059614260718,0)
(0.018181818181818077,1)
(0.0182648401826484,0)
(0.01984126984126977,0)
(0.022222222222222143,0)
(0.023693605972086917,1)
(0.023728813559322104,1)
(0.023728813559322104,1)
(0.023728813559322104,0)
(0.02499999999999991,1)
(0.025290498974709585,0)
(0.025290498974709585,0)
(0.027027027027026973,0)
(0.027450980392156765,1)
(0.027450980392156765,0)
(0.028094820017559252,0)
(0.028340080971659853,1)
(0.028919330289193246,1)
(0.030303030303030276,0)
(0.03214494447691418,1)
(0.032258064516129004,1)
(0.032258064516129004,1)
(0.032258064516129004,1)
(0.032738095238095344,1)
(0.03407984420642651,1)
(0.03442340791738374,1)
(0.03442340791738374,0)
(0.03442340791738374,0)
(0.03543098889476459,1)
(0.03599374021909241,1)
(0.03599374021909241,1)
(0.03838383838383841,1)
(0.03850931677018643,0)
(0.03850931677018643,1)
(0.03855855855855861,0)
(0.040000000000000036,1)
(0.041095890410958846,0)
(0.0412642669007901,0)
(0.042105263157894646,1)
(0.042105263157894646,0)
(0.042105263157894646,0)
(0.04428904428904423,1)
(0.04428904428904423,0)
(0.04480874316939887,0)
(0.045324153757888785,1)
(0.045324153757888785,0)
(0.04627450980392167,1)
(0.04627450980392167,1)
(0.04761904761904767,0)
(0.04761904761904767,1)
(0.04850964348334297,0)
(0.04850964348334297,1)
(0.048530416951469535,1)
(0.05129561078794298,0)
(0.05129561078794298,1)
(0.05138339920948609,1)
(0.05138339920948609,0)
(0.05138339920948609,1)
(0.05153153153153145,1)
(0.05153153153153145,1)
(0.05258799171842643,0)
(0.05258799171842643,0)
(0.05258799171842643,1)
(0.05258799171842643,1)
(0.05258799171842643,0)
(0.05384615384615388,0)
(0.056189640035118504,1)
(0.05723905723905731,1)
(0.058461538461538565,0)
(0.05882352941176472,1)
(0.05882352941176472,1)
(0.06034801925212885,0)
(0.062200956937799035,1)
(0.06416275430359941,0)
(0.06832298136645965,1)
(0.06832298136645965,1)
(0.07115384615384612,1)
(0.07115384615384612,1)
(0.0730282375851996,1)
(0.07457627118644061,1)
(0.07878787878787885,0)
(0.08333333333333326,1)
(0.08333333333333326,1)
(0.08333333333333326,1)
(0.08333333333333326,1)
(0.08333333333333326,1)
(0.08571428571428563,1)
(0.08571428571428563,1)
(0.08675799086757996,1)
(0.08771929824561409,1)
(0.08860759493670889,0)
(0.08937070333157049,1)
(0.09333333333333327,1)
(0.09333333333333327,0)
(0.09333333333333327,0)
(0.09604519774011289,1)
(0.09859154929577474,0)
(0.0990990990990992,0)
(0.10040160642570273,0)
(0.10476190476190483,1)
(0.10502283105022836,1)
(0.11158117398202005,1)
(0.11158117398202005,1)
(0.11163062536528345,0)
(0.11163062536528345,1)
(0.11163062536528345,1)
(0.11260504201680677,0)
(0.11260504201680677,1)
(0.11260504201680677,1)
(0.11384615384615393,1)
(0.11384615384615393,0)
(0.11384615384615393,1)
(0.11462450592885376,0)
(0.11578947368421044,1)
(0.11888111888111896,1)
(0.12786885245901636,1)
(0.1287878787878789,0)
(0.1287878787878789,1)
(0.1287878787878789,1)
(0.13590692755156009,1)
(0.13734658094681462,1)
(0.1384615384615384,1)
(0.1384615384615384,1)
(0.1457627118644067,1)
(0.1461187214611872,1)
(0.14782608695652177,1)
(0.14782608695652177,0)
(0.15294117647058814,0)
(0.1595959595959595,0)
(0.15961538461538471,1)
(0.15999999999999992,1)
(0.1619718309859155,0)
(0.16234796404019036,1)
(0.16234796404019036,1)
(0.16532258064516125,1)
(0.16532258064516125,1)
(0.17184265010351973,0)
(0.1729323308270676,0)
(0.1729323308270676,0)
(0.18128654970760238,0)
(0.18128654970760238,1)
(0.18128654970760238,1)
(0.18153846153846165,0)
(0.18181818181818188,0)
(0.18205128205128207,0)
(0.18205128205128207,1)
(0.1827027027027026,0)
(0.1909042834479111,0)
(0.19579193454120403,1)
(0.19617224880382778,0)
(0.20029618659755655,0)
(0.2045197740112994,1)
(0.20734126984126977,1)
(0.20934761441090566,0)
(0.20934761441090566,1)
(0.21727395411605932,1)
(0.22077922077922074,1)
(0.22157588577472231,1)
(0.22331002331002336,1)
(0.22331002331002336,1)
(0.2248447204968944,0)
(0.226936026936027,0)
(0.226936026936027,1)
(0.24,1)
(0.24242424242424243,0)
(0.24657534246575352,1)
(0.25383022774327113,0)
(0.26054054054054054,1)
(0.2615384615384615,1)
(0.2615384615384615,1)
(0.2615384615384615,1)
(0.2635885447106956,1)
(0.27231638418079096,1)
(0.27231638418079096,1)
(0.28248796741947424,0)
(0.2846153846153847,1)
(0.28671328671328666,1)
(0.28724440116845185,1)
(0.28724440116845185,0)
(0.28853754940711474,1)
(0.29824561403508776,0)
(0.29824561403508776,0)
(0.305050505050505,1)
(0.305050505050505,0)
(0.3055555555555556,1)
(0.3096045197740114,0)
(0.30980392156862746,1)
(0.31060606060606055,0)
(0.31060606060606055,1)
(0.31580655631288534,1)
(0.31677018633540377,1)
(0.3262032085561497,1)
(0.3262032085561497,0)
(0.3262823902696985,1)
(0.3262823902696985,1)
(0.33063427800269896,1)
(0.33333333333333326,1)
(0.33333333333333326,1)
(0.3407364114552893,0)
(0.3491525423728814,1)
(0.3538461538461539,0)
(0.37225636523266026,0)
(0.38636363636363646,1)
(0.3954116059379218,1)
(0.3954116059379218,1)
(0.3954116059379218,1)
(0.40639269406392686,0)
(0.4066631411951349,1)
(0.4076923076923078,1)
(0.41512605042016815,1)
(0.4385964912280702,1)
(0.4385964912280702,1)
(0.4490929285449834,1)
(0.4500264410364887,1)
(0.4500264410364887,1)
(0.4556451612903225,1)
(0.4584615384615385,1)
(0.4584615384615385,1)
(0.4584615384615385,1)
(0.4696969696969697,1)
(0.4813559322033898,1)
(0.49550502379693273,1)
(0.5256916996047432,1)
(0.5411605937921729,1)
(0.5411605937921729,1)
(0.5430988894764675,1)
(0.6023391812865497,1)
(0.6023391812865497,1)
(0.6023391812865497,1)
(0.6040404040404039,1)
(0.6210526315789473,1)
(0.622693096377307,1)
(0.6590909090909092,1)
(0.6684491978609626,0)
(0.6773109243697479,1)
(0.7046153846153846,0)
(0.7046153846153846,1)
(0.7153846153846153,1)
(0.7153846153846153,1)
(0.7468926553672317,0)
(0.7894736842105263,0)
(0.8461538461538463,1)
(0.8823529411764706,1)
(1.0,0)
(1.0,1)
(1.0,1)
    };
      
\end{axis}    
\end{tikzpicture}

\vspace{-0.1in}
\vspace{-0.5\baselineskip}
\caption{Accuracy of predictor as a function of agreement ($\kappa$), with logistic regression lines plotted. $R^2$ and p-values are listed below each title. Results are 0 or 1 but jitter is added to the y-axis for a more comprehensive illustration.}
\label{fig:pref_accuracy_agreementbin_corr}
\vspace{-0.5\baselineskip}
\end{figure}

\medskip
\noindent \textbf{Agreement Effect}
Here we analyze the effect of agreement on the accuracy of each predictor. We measured agreement among GuessTheKarma players by the Fleiss $\kappa$ score.
We plot the accuracy as a function of the agreement in Fig.~\ref{fig:pref_accuracy_agreementbin_corr} along with a logistic regression line and its associated coefficient of determination and p-value.
We find a weak, but statistically significant correlation; accuracy of the platform predictor as indicator of majority vote of the GuessTheKarma players increases from about 60\% in the low agreement range to about 75\% in the high agreement range.

\begin{figure}[t]
\centering 
\begin{tikzpicture}
\sffamily
\begin{axis}[ 
    title = {\small Reddit Vote-Scores \\\small $R^2=0.74$, $p=0.0063$},
    title style={align=center,yshift=-.1in},
    legend style={font=\scriptsize},
    width = 2in, height = 1.35in,
    enlarge y limits=0.01, 
    ylabel near ticks,
    xlabel near ticks,
    ylabel = {\small Accuracy},
    ymax=1, ymin=0.4,
    every tick label/.append style={font=\scriptsize},
    xmin=-1,xmax=20,
    domain=0:20,
    xlabel={ \small Subscribers (millions) },
    xlabel style = {yshift=0.05in},
    ] 
    \addplot+[no marks,solid,red]{-0.0119486004767*x +0.745037};
    \node[red] at (axis cs: 14,.49) {\tiny{$-0.01x+.75$}};
    \addplot[only marks,mark options={scale=.6}, red, fill=red!30!white, error bars/.cd, y dir=both, y explicit] coordinates {
		(0.03656, 0.68 ) +- (0, 0.133916988816 ) 
		(0.174044, 0.82 ) +- (0, 0.110293370493 ) 
		(0.247151, 0.770833333333 ) +- (0, 0.123332915968 ) 
		(5.064231, 0.64 ) +- (0, 0.137799444651 ) 
		(7.058654, 0.714285714286 ) +- (0, 0.131103572625 ) 
		(8.809353, 0.645833333333 ) +- (0, 0.140341819471 ) 
		(10.450075, 0.64 ) +- (0, 0.137799444651 ) 
		(18.399582, 0.530612244898 ) +- (0, 0.144832850594 ) 
    };
\end{axis}    
\end{tikzpicture}
\begin{tikzpicture}
\sffamily
\begin{axis}[ 
    title = {\small Imgur View-Counts \\\small $R^2=0.64$, $p=0.0165$},
    title style={align=center,yshift=-.1in}, 
    legend style={font=\scriptsize},
    width = 2in, height = 1.35in,
    enlarge y limits=0.01, 
    ylabel near ticks,
    xlabel near ticks,
    ymax=1, ymin=0.4,
    every tick label/.append style={font=\scriptsize}, 
    xmin=-1,xmax=20,
    domain=0:20,
    xlabel={ \small Subscribers (millions) },
    xlabel style = {yshift=0.05in},
    ] 
    \addplot+[no marks,solid,green]{-0.0111971831644*x +0.707818};
    \node[green] at (axis cs: 14,.47) {\tiny{$-0.01x+.71$}};
    \addplot[only marks,mark options={scale=.6}, green, fill=green!30!white, error bars/.cd, y dir=both, y explicit] coordinates {
		(0.03656, 0.64 ) +- (0, 0.137799444651 ) 
		(0.174044, 0.76 ) +- (0, 0.122607954468 ) 
		(0.247151, 0.729166666667 ) +- (0, 0.130402973178 ) 
		(5.064231, 0.68 ) +- (0, 0.133916988816 ) 
		(7.058654, 0.69387755102 ) +- (0, 0.133752397233 ) 
		(8.809353, 0.541666666667 ) +- (0, 0.146210919878 ) 
		(10.450075, 0.62 ) +- (0, 0.139345796711 ) 
		(18.399582, 0.510204081633 ) +- (0, 0.14507484383 ) 
     };
     
\end{axis}    
\end{tikzpicture}

\vspace{-0.15in}
\caption{Accuracy for each subreddit as a function of the number of subscribers. Linear regression lines are plotted with $R^2$ and p-values below the titles. Accuracy is negatively correlated with subreddit size ($p<0.05$). }
\label{fig:pref_accuracy_sr_size}
\end{figure}
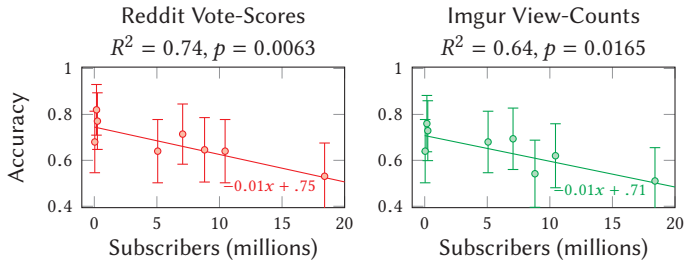

\medskip
\noindent \textbf{Subreddit Effect}
Next, we examine whether the accuracy of our predictors is affected by the choice of subreddit. 
Figure~\ref{fig:pref_accuracy_sr_size} shows that accuracy did vary by subreddit. It appears to be negatively correlated with size of the subreddits. The effect is statistically significant ($p<0.05$) but with so few subreddits we should be wary of reading too much into this correlation.

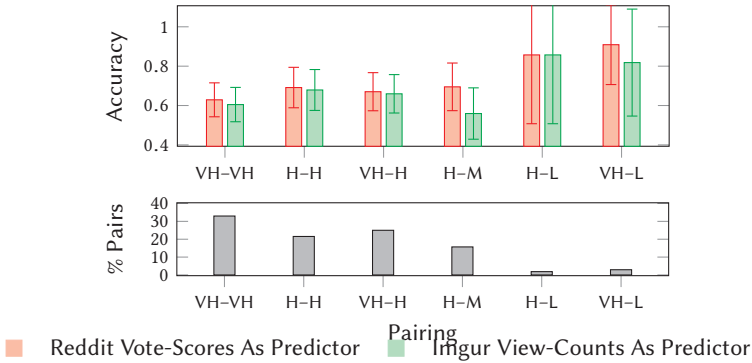
\begin{figure}[t]
\vspace{-0.25\baselineskip}
\small
\centering
\begin{tikzpicture}
\sffamily
\begin{axis}[
    ybar,
    xtick pos = left, 
    legend style={font=\scriptsize},
    width = 3.20in, height = 1.35in,
    enlarge y limits=0.01,
    enlarge x limits=0.12,   
    ylabel near ticks,
    xlabel near ticks,
    ymax=1.1, ymin=0.4,
    every tick label/.append style={font=\scriptsize}, 
    ylabel = {\small Accuracy},
        xticklabels={VH--VH, H--H, VH--H, H--M,H--L,VH--L,},
    xtick={1,2,3,4,5,6}, bar width=6pt,
    legend style={
    at={(0.05,0.95)},
    draw=none,
    anchor=north west}, legend cell align=left
    ] 
    \addplot[red, fill=red!30!white, error bars/.cd, y dir=both, y explicit] coordinates {
		(1, 0.629032258065 ) +- (0, 0.0862172229059 ) 
		(2, 0.691358024691 ) +- (0, 0.102778179916 ) 
		(3, 0.670212765957 ) +- (0, 0.0968094375418 ) 
		(4, 0.694915254237 ) +- (0, 0.12102217529 ) 
		(5, 0.857142857143 ) +- (0, 0.349558835542 ) 
		(6, 0.909090909091 ) +- (0, 0.202558077451 ) 
    };

    \addplot[green, fill=green!30!white, error bars/.cd, y dir=both, y explicit] coordinates {
		(1, 0.604838709677 ) +- (0, 0.0872562379485 ) 
		(2, 0.679012345679 ) +- (0, 0.103873537816 ) 
		(3, 0.659574468085 ) +- (0, 0.0975747414475 ) 
		(4, 0.559322033898 ) +- (0, 0.130490957491 ) 
		(5, 0.857142857143 ) +- (0, 0.349558835542 ) 
		(6, 0.818181818182 ) +- (0, 0.271760178344 ) 
    };

\end{axis}  
\end{tikzpicture}

\begin{tikzpicture}
\sffamily
\begin{axis}[
    ybar, 
    xtick pos = left,
    legend style={font=\scriptsize},
    width = 3.20in, height = 1.0in,
    enlarge y limits=0.01,
    enlarge x limits=0.12, 
    ylabel near ticks,
    xlabel near ticks,
    ymax=40, ymin=0, 
    every tick label/.append style={font=\scriptsize}, 
    xlabel = {\small Pairing},
    ylabel={\small \% Pairs}, ylabel style = {yshift=0.06in},
    xticklabels={VH--VH, H--H, VH--H, H--M,H--L,VH--L,},
    xtick={1,2,3,4,5,6},
    bar width=8pt, 
    ]
    
    \addplot[black, fill=black!30!white] coordinates { 
		(1, 32.9787  )  
		(2, 21.54255  ) 
		(3, 25  ) 
		(4, 15.69  ) 
		(5, 1.86170  ) 
		(6, 2.9255  ) 
    };
     
\end{axis}  
\end{tikzpicture}

\vspace{-1.0\baselineskip}

    \begin{tikzpicture} 
    \begin{axis}[%
    hide axis,
	width=2in, height=.7in,
    xmin=0,xmax=2,ymin=0,ymax=0.4,
    legend style={at={(1.5,0)},draw=none,fill=none, legend cell align=left,legend columns = -1, column sep = 3mm}
    ] 
    \addlegendimage{mark=square*,only marks,mark options={scale=1.5}, red!30!white}
    \addlegendentry{\small \sffamily Reddit Vote-Scores As Predictor};
    \addlegendimage{mark=square*,only marks,mark options={scale=1.5}, green!30!white}
    \addlegendentry{\small \sffamily Imgur View-Counts As Predictor};
    \end{axis}
    \end{tikzpicture}
\vspace{-0.1in}
\caption{
Accuracy for each pairing type ordered left to right by the size of score differentials. The distribution of image-pairs within each pairing type are plotted on bottom. } 
\label{fig:pref_accuracy_pairlogic}
\end{figure}

\medskip
\noindent \textbf{Image-pair Score Balance}
Here we compare platform accuracy with respect to the image-pair's score balance. We expect that if an image-pair is severely unbalanced (\eg, VH-L), then it is more likely that the groundtruth preference of GuessTheKarma players is aligned in favor of the VH post. Figure~\ref{fig:pref_accuracy_pairlogic} shows the accuracy of the platform predictors according to their pairing balance.  As expected, image-pairs with the largest imbalance resulted in significantly higher accuracy from the Reddit predictor ($p<0.05$) than the pairs where both images were very high-scoring.

Our primary concern with the previous image-pair balance analysis is that image scores within labels are not compared. So we instead compare our platform predictor as a function of the difference between the images' score-percentiles. For example, given a pair composed of images A and B, where A has a higher score than B; if A has a score that is higher than 95\% of the other images in the subreddit (\ie, a score-percentile of 95\%), and B has a score-percentile of 60\%, then the score-percentile $\Delta$ is 95\% - 60\% = 35\%. The score-percentile $\Delta$ is therefore a more fine-grained measure of the balance of an image-pair. 

\begin{figure}[t]
\small
\centering 
\begin{tikzpicture}
\sffamily
\begin{axis}[ 
    title = {Reddit Vote-Scores As Predictor\\$R^2=0.21$, $p=0.005$}, 
    title style={align=center,yshift=-.1in}, 
    legend style={font=\scriptsize},
    width = 2.75in, height = 1.35in, 
    ylabel near ticks,
    ylabel = {\small Accuracy},
    xlabel near ticks,
    ymax=1.2, ymin=-.2,
    every tick label/.append style={font=\scriptsize},
    xmin=-0.1,xmax=1.1,  
    xtick={0,0.2,0.4,0.6,0.8,1.0},
    xticklabels={0\%, 20\%, 40\%, 60\%, 80\%, 100\%},
    xlabel={\small Reddit Score-Percentile $\Delta$},
    xlabel style = {yshift=0.05in},
    ]  
    \addplot[no marks, solid, red, domain=-0.0:1]{(1/(1+exp(-(0.57772815+1.47762918*x))))};
    \node[red] at (axis cs: .85,.68) {\footnotesize{$1/e^{-(0.58+1.5x)}$}};
    \addplot[only marks,mark options={scale=.6}, red, fill=red!30!white, jitter=0.1 ] coordinates { 
	
	(0.36037966932,0.95)	(0.2538427221230001,0)	(0.012015177066000192,0.95)	(0.013685695503000028,0)	(0.198397737983,0.95)	(0.10701023081000004,0.95)	(0.10955966899000004,0.95)	(0.029610829102999836,0)	(0.03296801571399988,0)	(0.0137609767789999,0)	(0.03842676310999982,0.95)	(0.036677023070999966,0.95)	(0.02473546744099997,0)	(0.2673434856179999,0.95)	(0.07142857142900005,0.95)	(0.12884267631100002,0.95)	(0.007061502164000144,0.95)	(0.17107269954300008,0)	(0.14844941535300005,0)	(0.3432601935890001,0)	(0.05959302325599991,0.95)	(0.10055464834700012,0.95)	(0.024598914083999945,0.95)	(0.04256712508200011,0)	(0.10096025919500007,0)	(0.004761904762000002,0)	(0.033333333332999926,0.95)	(0.03749999999999998,0)	(0.03030303030300008,0)	(0.013096351730999987,0.95)	(0.0016175342329999287,0)	(0.3885585085919999,0.95)	(0.344466273739,0.95)	(0.30575964008799994,0.95)	(0.012442698101000071,0)	(0.01630206838699999,0.95)	(0.06818181818100011,0.95)	(0.024472573839000037,0)	(0.06666666666599996,0)	(0.00775193798399998,0)	(0.113095238,0.95)	(0.07317927099999999,0.95)	(0.009883198999999898,0)	(0.125,0.95)	(0.125,0.95)	(0.033613444999999964,0)	(0.8235294119999998,0.95)	(0.4689542489999999,0.95)	(0.05000000000000005,0)	(0.04545454500000001,0.95)	(0.03773584900000004,0.95)	(0.07534246599999994,0.95)	(0.012987012999999936,0.95)	(0.094339622,0.95)	(0.07297447200000008,0.95)	(0.047619047999999935,0.95)	(0.0,0)	(0.0,0.95)	(0.2590970349999999,0)	(0.026744186047000015,0.95)	(0.023255813953000026,0.95)	(0.807692307692,0.95)	(0.0,0.95)	(0.15476190476099996,0)	(0.261904761904,0.95)	(0.11666666699999995,0.95)	(0.08333333399999998,0.95)	(0.08333333300000001,0.95)	(0.335443038,0)	(0.156924795,0)	(0.03501561700000011,0.95)	(0.6910409240000001,0.95)	(0.043290042999999945,0.95)	(0.05869603000000001,0)	(0.047619047999999935,0.95)	(0.05000000000000005,0.95)	(0.559322034,0.95)	(0.1593026030000001,0.95)	(0.3164556960000001,0.95)	(0.11392405100000015,0)	(0.482180164,0.95)	(0.08860759500000004,0.95)	(0.25,0.95)	(0.0,0)	(0.08333333300000001,0.95)	(0.153846154,0.95)	(0.028571429000000093,0)	(0.0036231890000000266,0.95)	(0.05476190399999992,0)	(0.09959047,0.95)	(0.13235294200000006,0.95)	(0.13235294100000014,0.95)	(0.03076923099999995,0.95)	(0.07692307700000013,0.95)	(0.14505494499999996,0.95)	(0.0769230769999999,0)	(0.096153846,0.95)	(0.166666667,0.95)	(0.05000000000000005,0.95)	(0.028571429000000093,0)	(0.157692308,0.95)	(0.09999999999999998,0.95)	(0.2,0.95)	(0.2,0.95)	(0.036956521999999985,0.95)	(0.1499999999999999,0.95)	(0.036363636362999885,0.95)	(0.12467532467499988,0.95)	(0.12991944764099994,0.95)	(0.008830022074999944,0.95)	(0.05151515151500008,0)	(0.08896103896100005,0)	(0.16363636363599987,0.95)	(0.028329809724999988,0.95)	(0.017485984484000072,0.95)	(0.06578947368400012,0.95)	(0.11508771929900008,0.95)	(0.06849315,0.95)	(0.21428571428600007,0.95)	(0.10176565008100004,0.95)	(0.014697569247999918,0.95)	(0.1041426927500001,0.95)	(0.07348798016800008,0.95)	(0.07633018398800008,0.95)	(0.13062318424600006,0)	(0.22561271717199996,0)	(0.03475402117899995,0.95)	(0.159015359656,0.95)	(0.022853747777999974,0.95)	(0.002697760943000005,0)	(0.03448275862100003,0.95)	(0.013513513513999986,0.95)	(0.18014022888300008,0)	(0.013108614231999938,0)	(0.014243001749000063,0)	(0.016853932584999898,0)	(0.0024968789020000104,0.95)	(0.077351010316,0.95)	(0.23595505617900006,0)	(0.0927518427520001,0)	(0.14285714299999994,0.95)	(0.11688311699999998,0)	(0.196969697,0.95)	(0.0625,0)	(0.01793032786899995,0.95)	(0.07685810810799998,0.95)	(0.10267857142900004,0.95)	(0.184782609,0)	(0.2307692309999999,0)	(0.08695652200000004,0.95)	(0.664961637,0.95)	(0.012987012999999936,0.95)	(0.07388663967600008,0)	(0.125,0.95)	(0.06666666666700005,0.95)	(0.04523809523799993,0.95)	(0.05000000000000005,0)	(0.1428571428569999,0.95)	(0.05000000000000005,0)	(0.032258064516,0)	(0.05000000000000005,0.95)	(0.032258064516,0.95)	(0.0,0.95)	(0.04545454500000001,0.95)	(0.04040403999999997,0.95)	(0.014285714000000003,0.95)	(0.7424242420000001,0.95)	(0.08333333300000001,0)	(0.028571429000000093,0)	(0.03846153846199996,0.95)	(0.02097902097899995,0)	(0.0,0)	(0.05714285699999987,0.95)	(0.330952381,0.95)	(0.061663652999999985,0.95)	(0.014285715000000088,0.95)	(0.009164419999999951,0)	(0.12770562799999996,0.95)	(0.09877253599999992,0.95)	(0.181818182,0.95)	(0.235294118,0)	(0.07534246599999994,0)	(0.2705244130000001,0)	(0.752704577,0.95)	(0.02611832599999997,0.95)	(0.10714285714300008,0)	(0.2142857142850001,0)	(0.03666666699999999,0.95)	(0.6333333329999999,0.95)	(0.35,0.95)	(0.21739130400000006,0)	(0.559322034,0.95)	(0.0418743760000001,0.95)	(0.5423728809999999,0.95)	(0.853050847,0.95)	(0.036923076923000016,0)	(0.040000000000000036,0)	(0.04761904761899993,0.95)	(0.00541711809299994,0.95)	(0.023469410862999918,0)	(0.02429945776699993,0)	(0.08113425442699984,0.95)	(0.03830581295399993,0.95)	(0.10996749729200006,0.95)	(0.0025265037929999012,0)	(0.12707762561699998,0)	(0.021434231031000058,0)	(0.033684210526000014,0)	(0.005459085670999953,0)	(0.05911623884999995,0.95)	(0.15429098314500012,0.95)	(0.14511041009400005,0.95)	(0.2260778128280001,0)	(0.5272792732910001,0)	(0.10351089588399986,0)	(0.01163639789899995,0)	(0.046732995379,0.95)	(0.02650316946400011,0.95)	(0.04678906716899989,0.95)	(0.0163438256659999,0)	(0.016041162227000028,0.95)	(0.006203473946000071,0.95)	(0.03448275862100003,0.95)	(0.03846153846200007,0)	(0.2274358974360001,0.95)	(0.01780626780600003,0.95)	(0.013380135135999892,0.95)	(0.0057433954019999725,0)	(0.03953370501800002,0.95)	(0.028571428570999924,0)	(0.20559246954600008,0.95)	(0.043140485588999966,0.95)	(0.028838750131999987,0.95)	(0.00043815222499998807,0.95)	(0.1945440198120002,0.95)	(0.12949822605200012,0)	(0.214285714285,0.95)	(0.17857142857099995,0)	(0.042207792207,0)	(0.03515132408499999,0.95)	(0.032459016393000084,0.95)	(0.03278688524499995,0.95)	(0.15189873417700006,0.95)	(0.11392405063300015,0)	(0.1037548001709998,0.95)	(0.22151898734199996,0.95)	(0.06329113924100005,0)	(0.009367828931999944,0.95)	(0.037068183305000064,0)	(0.00841750841799993,0.95)	(0.053571428570999946,0.95)	(0.023255813952999915,0)	(0.09302325581399996,0)	(0.004617304606000072,0)	(0.047156726768999964,0)	(0.32675725111,0)	(0.02502207830400005,0)	(0.058139534884000026,0.95)	(0.014244186046999953,0.95)	(0.06906431853700001,0)	(0.30258173042400005,0)	(0.08095378274999998,0.95)	(0.10465116279100006,0.95)	(0.11545103709800007,0.95)	(0.062178828366000076,0.95)	(0.13980022986999985,0.95)	(0.005353430836000039,0)	(0.095530726257,0.95)	(0.570238095238,0.95)	(0.013095238094999906,0)	(0.025000000000000026,0.95)	(0.1511627906980001,0.95)	(0.025000000000000026,0.95)	(0.0,0.95)	(0.04166666666700003,0.95)	(0.012499999999999956,0.95)	(0.09125116975299996,0.95)	(0.0,0.95)	(0.023255813953000026,0.95)	(0.01818181818200004,0)	(0.098360655738,0.95)	(0.2493271348180001,0.95)	(0.04876718448400008,0.95)	(0.13121316643100012,0)	(0.16307106859099996,0)	(0.0808625336929999,0.95)	(0.023255813953000026,0.95)	(0.03030303030300008,0.95)	(0.006058739877000008,0.95)	(0.011478368129999916,0.95)	(0.026280323449999976,0.95)	(0.01969696969699997,0.95)	(0.015105164620999978,0.95)	(0.05000000000000005,0.95)	(0.16666666666700006,0.95)	(0.03846153846200007,0.95)	(0.015151515150999884,0)	(0.06210021322000015,0.95)	(0.1641791044780001,0.95)	(0.34294436906400017,0.95)	(0.1428571428569999,0.95)	(0.230769230769,0.95)	(0.040000000000000036,0.95)	(0.0,0)	(0.0,0)	(0.0,0.95)	(0.0,0.95)	(0.04761904761899993,0.95)	(0.13636363636399995,0)	(0.023708287323999988,0)	(0.17198905109499985,0.95)	(0.17198905109499985,0.95)	(0.09279918864000014,0.95)	(0.01818181818200004,0.95)	(0.028877005346999995,0.95)	(0.06827242524900012,0)	(0.02228810879200005,0)	(0.09534671532899984,0.95)	(0.07604348471299982,0.95)	(0.044628042870999966,0)	(0.04426213412399993,0)	(0.018540137463999958,0)	(0.13991163475699994,0.95)	(0.05356206784600004,0)	(0.1586186300150001,0)	(0.02285078487899983,0.95)	(0.274585392708,0)	(0.2983234714009999,0.95)	(0.13547340665500007,0.95)	(0.06912108310800003,0.95)	(0.10397946084800014,0.95)	(0.022383589853999908,0.95)	(0.011217736516999866,0.95)	(0.004467683726000038,0.95)	(0.2455903297669999,0)	(0.3105515587529999,0.95)	(0.18032786885199992,0.95)	(0.10810810810799998,0.95)	(0.159090909091,0.95)	(0.13723776223700002,0)	(0.11363636363600005,0)	(0.011549925484000023,0.95)	(0.363636363637,0.95)	(0.0051354577580000615,0)	(0.02578715924399999,0)	(0.11907066795699996,0.95)	(0.185499316006,0.95)	(0.012584704743999977,0.95)	(0.0221511183580001,0)	(0.12875121006800014,0.95)	(0.04646660212999998,0.95)	(0.06105577763000002,0.95)	(0.08644221059400015,0.95)	(0.058526675954000074,0.95)	(0.13103953147900016,0.95)	(0.00962861072900012,0.95)	(0.10402405000000003,0.95)	(0.035505124450999966,0)	(0.13396778916599986,0.95)	(0.277086383602,0.95)	(0.01891906341299998,0.95)	(0.240077444337,0.95)	(0.0011261261259999378,0.95)	(0.028153153153000043,0.95)	(0.03194103194099995,0.95)	(0.638888888889,0.95)	(0.04918032786899995,0.95)	(0.032258064516,0.95)	(0.8793933558020001,0.95)	(0.01612903225800011,0.95)	(0.0501644736839999,0.95)	(0.02141203703700001,0)	(0.02491554054100009,0)	(0.03846153846199996,0.95)	(0.15384615384599998,0.95)	(0.019230769230999974,0)	(0.3333333333330001,0.95)	(0.04761904761899993,0)	(0.06593406593399986,0.95)	(0.17307692307699996,0.95)	(0.038461538460999985,0.95)	(0.04545454545499994,0.95)	(0.0,0)	(0.040000000000000036,0.95)	(0.550909090909,0.95)	(0.2467532467530001,0.95)	(0.27272727272700004,0.95)	(0.02164502164500004,0.95)
    };
    
\end{axis}    
\end{tikzpicture} 
\begin{tikzpicture}
\sffamily
\begin{axis}[ 
    title = {Imgur View-Counts As Predictor \\$R^2=0.17$, $p=0.012$}, 
    title style={align=center,yshift=-.1in}, 
    legend style={font=\scriptsize},
    width = 2.75in, height = 1.35in,  
    ylabel near ticks,
    xlabel near ticks,
    ymax=1.2, ymin=-.2, 
    every tick label/.append style={font=\scriptsize},
    xmin=-0.1,xmax=1.1,   
    xtick={0,0.2,0.4,0.6,0.8,1.0},
    xticklabels={0\%, 20\%, 40\%, 60\%, 80\%, 100\%}, 
    xlabel={\small Imgur Score-Percentile $\Delta$},
    xlabel style = {yshift=0.05in},
    ] 
     
    \addplot[no marks, solid, green, domain=-0.0:1]{(1/(1+exp(-(0.50325739+1.30108635*x))))};
    \node[green] at (axis cs: .85,.68) {\footnotesize{$1/e^{-(0.50+1.3x)}$}};
    \addplot[only marks,mark options={scale=.6}, green, fill=green!30!white, jitter=0.1  ] coordinates { 
	
	(0.36037966932,0.95)	(0.2538427221230001,0.95)	(0.012015177066000192,0.95)	(0.013685695503000028,0.95)	(0.198397737983,0.95)	(0.10701023081000004,0)	(0.10955966899000004,0.95)	(0.029610829102999836,0)	(0.03296801571399988,0)	(0.0137609767789999,0)	(0.03842676310999982,0.95)	(0.036677023070999966,0)	(0.02473546744099997,0)	(0.2673434856179999,0.95)	(0.07142857142900005,0)	(0.12884267631100002,0.95)	(0.007061502164000144,0)	(0.17107269954300008,0.95)	(0.14844941535300005,0)	(0.3432601935890001,0)	(0.05959302325599991,0.95)	(0.10055464834700012,0)	(0.024598914083999945,0.95)	(0.04256712508200011,0)	(0.10096025919500007,0)	(0.004761904762000002,0.95)	(0.033333333332999926,0.95)	(0.03749999999999998,0)	(0.03030303030300008,0.95)	(0.013096351730999987,0.95)	(0.0016175342329999287,0)	(0.3885585085919999,0.95)	(0.344466273739,0.95)	(0.30575964008799994,0.95)	(0.012442698101000071,0)	(0.01630206838699999,0.95)	(0.06818181818100011,0.95)	(0.024472573839000037,0.95)	(0.06666666666599996,0.95)	(0.00775193798399998,0.95)	(0.113095238,0.95)	(0.07317927099999999,0.95)	(0.009883198999999898,0)	(0.125,0.95)	(0.125,0.95)	(0.033613444999999964,0)	(0.8235294119999998,0.95)	(0.4689542489999999,0.95)	(0.05000000000000005,0)	(0.04545454500000001,0.95)	(0.03773584900000004,0.95)	(0.07534246599999994,0.95)	(0.012987012999999936,0.95)	(0.094339622,0)	(0.07297447200000008,0.95)	(0.047619047999999935,0.95)	(0.0,0)	(0.0,0.95)	(0.2590970349999999,0)	(0.026744186047000015,0.95)	(0.023255813953000026,0.95)	(0.807692307692,0.95)	(0.0,0.95)	(0.15476190476099996,0.95)	(0.261904761904,0)	(0.11666666699999995,0.95)	(0.08333333399999998,0.95)	(0.08333333300000001,0)	(0.335443038,0)	(0.156924795,0)	(0.03501561700000011,0.95)	(0.6910409240000001,0.95)	(0.043290042999999945,0)	(0.05869603000000001,0.95)	(0.047619047999999935,0.95)	(0.05000000000000005,0.95)	(0.559322034,0.95)	(0.1593026030000001,0)	(0.3164556960000001,0.95)	(0.11392405100000015,0)	(0.482180164,0.95)	(0.08860759500000004,0.95)	(0.25,0.95)	(0.0,0.95)	(0.08333333300000001,0.95)	(0.153846154,0.95)	(0.028571429000000093,0)	(0.0036231890000000266,0.95)	(0.05476190399999992,0)	(0.09959047,0.95)	(0.13235294200000006,0.95)	(0.13235294100000014,0.95)	(0.03076923099999995,0.95)	(0.07692307700000013,0.95)	(0.14505494499999996,0)	(0.0769230769999999,0.95)	(0.096153846,0.95)	(0.166666667,0.95)	(0.05000000000000005,0.95)	(0.028571429000000093,0)	(0.157692308,0.95)	(0.09999999999999998,0)	(0.2,0.95)	(0.2,0.95)	(0.036956521999999985,0)	(0.1499999999999999,0.95)	(0.036363636362999885,0)	(0.12467532467499988,0)	(0.12991944764099994,0.95)	(0.008830022074999944,0.95)	(0.05151515151500008,0)	(0.08896103896100005,0)	(0.16363636363599987,0.95)	(0.028329809724999988,0.95)	(0.017485984484000072,0.95)	(0.06578947368400012,0.95)	(0.11508771929900008,0.95)	(0.06849315,0.95)	(0.21428571428600007,0.95)	(0.10176565008100004,0.95)	(0.014697569247999918,0.95)	(0.1041426927500001,0.95)	(0.07348798016800008,0.95)	(0.07633018398800008,0.95)	(0.13062318424600006,0)	(0.22561271717199996,0.95)	(0.03475402117899995,0.95)	(0.159015359656,0.95)	(0.022853747777999974,0.95)	(0.002697760943000005,0)	(0.03448275862100003,0.95)	(0.013513513513999986,0.95)	(0.18014022888300008,0)	(0.013108614231999938,0)	(0.014243001749000063,0)	(0.016853932584999898,0)	(0.0024968789020000104,0.95)	(0.077351010316,0.95)	(0.23595505617900006,0.95)	(0.0927518427520001,0)	(0.14285714299999994,0.95)	(0.11688311699999998,0)	(0.196969697,0.95)	(0.0625,0)	(0.01793032786899995,0.95)	(0.07685810810799998,0.95)	(0.10267857142900004,0.95)	(0.184782609,0.95)	(0.2307692309999999,0)	(0.08695652200000004,0.95)	(0.664961637,0.95)	(0.012987012999999936,0.95)	(0.07388663967600008,0)	(0.125,0.95)	(0.06666666666700005,0.95)	(0.04523809523799993,0)	(0.05000000000000005,0)	(0.1428571428569999,0.95)	(0.05000000000000005,0)	(0.032258064516,0)	(0.05000000000000005,0.95)	(0.032258064516,0.95)	(0.0,0.95)	(0.04545454500000001,0.95)	(0.04040403999999997,0.95)	(0.014285714000000003,0.95)	(0.7424242420000001,0.95)	(0.08333333300000001,0)	(0.028571429000000093,0)	(0.03846153846199996,0.95)	(0.02097902097899995,0)	(0.0,0)	(0.05714285699999987,0.95)	(0.330952381,0.95)	(0.061663652999999985,0.95)	(0.014285715000000088,0.95)	(0.009164419999999951,0)	(0.12770562799999996,0.95)	(0.09877253599999992,0.95)	(0.181818182,0.95)	(0.235294118,0)	(0.07534246599999994,0)	(0.2705244130000001,0)	(0.752704577,0.95)	(0.02611832599999997,0.95)	(0.10714285714300008,0)	(0.2142857142850001,0)	(0.03666666699999999,0.95)	(0.6333333329999999,0.95)	(0.35,0.95)	(0.21739130400000006,0)	(0.559322034,0.95)	(0.0418743760000001,0.95)	(0.5423728809999999,0.95)	(0.853050847,0.95)	(0.036923076923000016,0)	(0.040000000000000036,0)	(0.04761904761899993,0.95)	(0.00541711809299994,0.95)	(0.023469410862999918,0)	(0.02429945776699993,0)	(0.08113425442699984,0.95)	(0.03830581295399993,0.95)	(0.10996749729200006,0.95)	(0.0025265037929999012,0)	(0.12707762561699998,0)	(0.021434231031000058,0)	(0.033684210526000014,0)	(0.005459085670999953,0)	(0.05911623884999995,0.95)	(0.15429098314500012,0.95)	(0.14511041009400005,0.95)	(0.2260778128280001,0)	(0.5272792732910001,0)	(0.10351089588399986,0)	(0.01163639789899995,0)	(0.046732995379,0.95)	(0.02650316946400011,0.95)	(0.04678906716899989,0.95)	(0.0163438256659999,0)	(0.016041162227000028,0)	(0.006203473946000071,0.95)	(0.03448275862100003,0.95)	(0.03846153846200007,0)	(0.2274358974360001,0.95)	(0.01780626780600003,0.95)	(0.013380135135999892,0.95)	(0.0057433954019999725,0)	(0.03953370501800002,0.95)	(0.028571428570999924,0)	(0.20559246954600008,0.95)	(0.043140485588999966,0.95)	(0.028838750131999987,0.95)	(0.00043815222499998807,0)	(0.1945440198120002,0.95)	(0.12949822605200012,0)	(0.214285714285,0.95)	(0.17857142857099995,0)	(0.042207792207,0)	(0.03515132408499999,0.95)	(0.032459016393000084,0.95)	(0.03278688524499995,0.95)	(0.15189873417700006,0.95)	(0.11392405063300015,0)	(0.1037548001709998,0.95)	(0.22151898734199996,0.95)	(0.06329113924100005,0)	(0.009367828931999944,0.95)	(0.037068183305000064,0)	(0.00841750841799993,0.95)	(0.053571428570999946,0)	(0.023255813952999915,0)	(0.09302325581399996,0.95)	(0.004617304606000072,0)	(0.047156726768999964,0)	(0.32675725111,0)	(0.02502207830400005,0)	(0.058139534884000026,0)	(0.014244186046999953,0.95)	(0.06906431853700001,0)	(0.30258173042400005,0)	(0.08095378274999998,0.95)	(0.10465116279100006,0.95)	(0.11545103709800007,0.95)	(0.062178828366000076,0.95)	(0.13980022986999985,0.95)	(0.005353430836000039,0)	(0.095530726257,0.95)	(0.570238095238,0.95)	(0.013095238094999906,0)	(0.025000000000000026,0)	(0.1511627906980001,0.95)	(0.025000000000000026,0.95)	(0.0,0.95)	(0.04166666666700003,0.95)	(0.012499999999999956,0.95)	(0.09125116975299996,0.95)	(0.0,0.95)	(0.023255813953000026,0.95)	(0.01818181818200004,0)	(0.098360655738,0.95)	(0.2493271348180001,0.95)	(0.04876718448400008,0.95)	(0.13121316643100012,0)	(0.16307106859099996,0)	(0.0808625336929999,0.95)	(0.023255813953000026,0.95)	(0.03030303030300008,0.95)	(0.006058739877000008,0.95)	(0.011478368129999916,0.95)	(0.026280323449999976,0.95)	(0.01969696969699997,0.95)	(0.015105164620999978,0)	(0.05000000000000005,0.95)	(0.16666666666700006,0.95)	(0.03846153846200007,0.95)	(0.015151515150999884,0)	(0.06210021322000015,0.95)	(0.1641791044780001,0.95)	(0.34294436906400017,0)	(0.1428571428569999,0.95)	(0.230769230769,0.95)	(0.040000000000000036,0)	(0.0,0)	(0.0,0)	(0.0,0.95)	(0.0,0.95)	(0.04761904761899993,0.95)	(0.13636363636399995,0)	(0.023708287323999988,0.95)	(0.17198905109499985,0.95)	(0.17198905109499985,0.95)	(0.09279918864000014,0.95)	(0.01818181818200004,0.95)	(0.028877005346999995,0.95)	(0.06827242524900012,0)	(0.02228810879200005,0)	(0.09534671532899984,0.95)	(0.07604348471299982,0.95)	(0.044628042870999966,0)	(0.04426213412399993,0)	(0.018540137463999958,0)	(0.13991163475699994,0.95)	(0.05356206784600004,0.95)	(0.1586186300150001,0.95)	(0.02285078487899983,0.95)	(0.274585392708,0.95)	(0.2983234714009999,0.95)	(0.13547340665500007,0.95)	(0.06912108310800003,0.95)	(0.10397946084800014,0.95)	(0.022383589853999908,0.95)	(0.011217736516999866,0.95)	(0.004467683726000038,0)	(0.2455903297669999,0)	(0.3105515587529999,0.95)	(0.18032786885199992,0.95)	(0.10810810810799998,0.95)	(0.159090909091,0.95)	(0.13723776223700002,0)	(0.11363636363600005,0)	(0.011549925484000023,0.95)	(0.363636363637,0)	(0.0051354577580000615,0)	(0.02578715924399999,0)	(0.11907066795699996,0.95)	(0.185499316006,0.95)	(0.012584704743999977,0.95)	(0.0221511183580001,0)	(0.12875121006800014,0.95)	(0.04646660212999998,0.95)	(0.06105577763000002,0.95)	(0.08644221059400015,0.95)	(0.058526675954000074,0.95)	(0.13103953147900016,0.95)	(0.00962861072900012,0.95)	(0.10402405000000003,0.95)	(0.035505124450999966,0)	(0.13396778916599986,0)	(0.277086383602,0)	(0.01891906341299998,0.95)	(0.240077444337,0.95)	(0.0011261261259999378,0.95)	(0.028153153153000043,0.95)	(0.03194103194099995,0.95)	(0.638888888889,0.95)	(0.04918032786899995,0.95)	(0.032258064516,0.95)	(0.8793933558020001,0.95)	(0.01612903225800011,0.95)	(0.0501644736839999,0.95)	(0.02141203703700001,0)	(0.02491554054100009,0)	(0.03846153846199996,0.95)	(0.15384615384599998,0.95)	(0.019230769230999974,0.95)	(0.3333333333330001,0.95)	(0.04761904761899993,0)	(0.06593406593399986,0.95)	(0.17307692307699996,0.95)	(0.038461538460999985,0.95)	(0.04545454545499994,0.95)	(0.0,0.95)	(0.040000000000000036,0)	(0.550909090909,0)	(0.2467532467530001,0.95)	(0.27272727272700004,0.95)	(0.02164502164500004,0.95)

    }; 
\end{axis}    
\end{tikzpicture}
  
\vspace{-0.75\baselineskip}
\caption{Correctness of prediction for each image-pair plotted by the Reddit score-percentile difference ($\Delta$) with logistic regression lines plotted. Accuracy is weakly, positively correlated with Reddit score-percentile difference. R$^2$ and p-values are listed below each title. Results are 0 or 1 but jitter is added to the y-axis for a more comprehensive illustration.}  
\label{fig:pref_accuracy_scorediff_corr}
\end{figure}
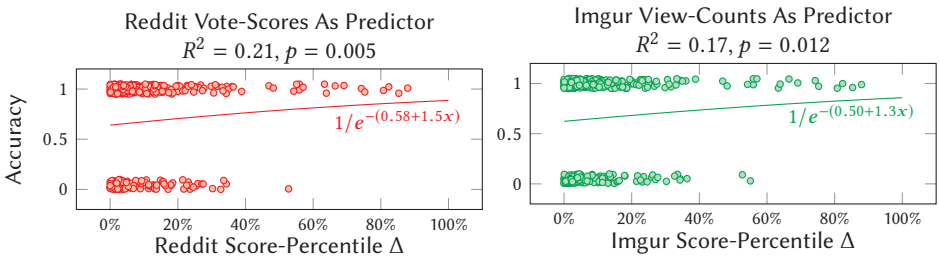

Figure~\ref{fig:pref_accuracy_scorediff_corr} shows that the score-percentile $\Delta$ of an image pair is significantly ($p<0.05$) correlated with accuracy for both predictors. This indicates, as expected, that image-pairs that received similar amounts of attention are more difficult to predict than unbalanced image-pairs. Further, we see that once the balance ratio reaches a threshold of 60\%, the platform predictors were able to accurately predict the images that the majority of our respondents preferred.

The image-pair balance results show that Reddit and to a lesser extent Imgur are able to accurately make binary predictions of user preference, but only when the image-pairs are severely unbalanced. It is important to note that most of the image-pairs are concentrated within the lower range of score-percentile $\Delta$, \ie, when the score-percentile $\Delta$ is less than 20\%. Overall, the predictive ability of these social platforms seems to be limited to only a few rare instances.

\subsection{Individual Player Ratings}

The previous analysis and results determined how accurately Reddit vote-scores and Imgur view-counts predict the majority preferences of GuessTheKarma players. The majority preferences were used as the groundtruth because they were collected without social influence and algorithmic-bias effects that are often found in social media systems.
Next we analyzed how well individual players predict the platform outcome. 
The results of this task were only slightly better than random guess: personal preference accuracy was 54.0\% with a 95\% confidence interval of $\pm$ 0.8\% and the player prediction accuracy was 60.6\% $\pm$ 0.6\%. We again emphasize that the player accuracy is measured differently than in the previous groundtruth analysis. Here, we measure the player accuracy as their individual ability to predict the winning image from Reddit scores or Imgur view-counts. 

\medskip
\noindent \textbf{Expertise Effect}
Do those who use Reddit frequently perform better than those who use Reddit casually or not at all? To answer this question we performed a battery of correlation and statistical tests comparing the accuracies of players grouped by their questionnaire responses. We used the one-tailed two valued $t$-test; for each question, the null hypothesis is that players who self report that they do not use Reddit have the same accuracy as those who answer otherwise. For example, we compare the accuracy distributions of players who self-reported heavy Reddit usage to those that do not use Reddit at all, and those who use Reddit casually to those that do not use Reddit at all, and likewise for the four other questions. We also performed the one-tailed two valued $t$-test where, for each question, the alternate hypothesis was that players who indicated heavy use had higher accuracy as players that indicated casual use.

There are three responses for each of five questions, resulting in dozens of statistical comparisons. Here we also employed Bonferroni correction to mitigate problems that arise when performing multiple statistical comparisons. No statistically significant correlations were found. 
In particular no difference was observed between the accuracy of users who were active within a subreddit (even a small subreddit) and non-users with mean accuracies of $62.8\% \pm 4.2$ and $63.5\% \pm 2.6$, respectively. 

Next we ask: do self-reported {\em powerusers}, \ie, those that indicated frequent Reddit use on all survey questions, perform better than non-powerusers? To answer this question we again use the one-tailed two value $t$-test with Bonferroni correction to correct for the multiple tests. We considered two null hypotheses: 1) that powerusers have the same accuracy as non-powerusers, and 2) that powerusers have the same accuracy as those players who do not use Reddit, \ie, players who answered ``Don't use Reddit'' to all questions. We also repeat this analysis for powerusers that answered Yes or No to the subreddit-use survey question.

Of the players who completed the survey, only 174 qualified to be called powerusers. Within this set of powerusers, 60 also reported that they use the subreddit from which their image-pairs were drawn. No statistically significant differences were found. Overall, we see no evidence of an expertise effect, lending confidence to our inclusion of raters who were unfamiliar with the particular subreddits.

\medskip
\noindent \textbf{Effort Effect}
Finally, we consider response times (delays between when a player is shown an image-pair and when they made a selection) as a proxy for player effort. We compare the response time distributions for correct and incorrect judgments to determine if incorrect judgments are correlated with player effort as measured by player response time.

\begin{figure}
    \centering
    \begin{tikzpicture}
\sffamily
\begin{axis}[
    title =,
    legend style={font=\scriptsize, draw= none, fill = none},
    every tick label/.append style={font=\scriptsize},
    width = 3.4in, height = 1.5in,
    xlabel = {\small Response Time}, 
    enlarge y limits=0.00, 
    ylabel near ticks,
    ymin=0, ymax =2000,
    xlabel near ticks, 
    y label style={at={(axis description cs:-0.08,.5)}}, 
    ]
  \addplot[blue, fill = blue!30!white, opacity=0.99] coordinates {
	(0.0,9)	(1.0,106)	(2.0,698)	(3.0,1319)	(4.0,1904)	(5.0,2012)	(6.0,1887)	(7.0,1634)	(8.0,1528)	(9.0,1231)	(10.0,1074)	(11.0,863)	(12.0,741)	(13.0,615)	(14.0,539)	(15.0,412)	(16.0,403)	(17.0,328)	(18.0,262)	(19.0,215)	(20.0,185)	(21.0,174)	(22.0,128)	(23.0,128)	(24.0,117)	(25.0,103)	(26.0,82)	(27.0,79)	(28.0,67)	(29.0,63)
  }; 
   \addplot[red, fill = red!30!white, opacity=0.99,] coordinates { 
	(0.0,6)	(1.0,81)	(2.0,399)	(3.0,734)	(4.0,972)	(5.0,1152)	(6.0,1168)	(7.0,1083)	(8.0,999)	(9.0,782)	(10.0,759)	(11.0,662)	(12.0,540)	(13.0,432)	(14.0,365)	(15.0,288)	(16.0,242)	(17.0,216)	(18.0,183)	(19.0,192)	(20.0,134)	(21.0,124)	(22.0,104)	(23.0,93)	(24.0,85)	(25.0,74)	(26.0,72)	(27.0,65)	(28.0,64)	(29.0,75)
  };
  \legend{Correct,Incorrect}
\end{axis}
\end{tikzpicture}
    \vspace{-1\baselineskip}
    \caption{Response time (seconds) for correct and incorrect player predictions. Distributions are not significantly different ($p=0.419$).}
    \label{fig:response_time}
    \vspace{-1\baselineskip}
\end{figure}
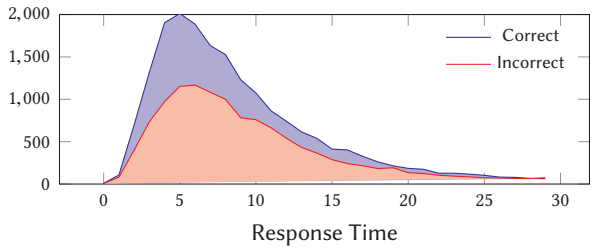

Using the two valued two-tailed $t$-test, we found that the response time distributions were similar for correct and incorrect predictions ($p=0.419$ with means of 22 and 26 seconds, respectively). Moreover, player response times, organized into one-second-sized bins and shown in  Fig.~\ref{fig:response_time} were not significantly correlated with prediction accuracy ($R^2=0.18$, $p=0.572$). Overall we find that there is little effect of player effort on prediction accuracy.

\section{Discussion}

Several methods of analysis were performed.  Altogether, the totality of the results leads us to conclude that, except in specific circumstances where image-pairs are highly unbalanced and players are in strong agreement, Reddit vote-scores and Imgur view-counts are relatively poor predictors of user preference. 

Our results are consistent with the finding of Salganik, Dodds, and Watts, in their study of artificial cultural markets~\cite{salganik2006experimental}. They simulated a music marketplace with social influence feedback loops; songs were shown in sorted order based on previous downloads. They ran eight versions of the marketplace, and one with random ordering to avoid social influence and thus measure true population preference. They found that the truly most preferred songs in the independent information condition rarely fell to the bottom and that the truly least preferred songs rarely stayed at the top but beyond that the ordering in any of the marketplaces was unpredictable from the true popularity. This matches our finding that GuessTheKarma players generally preferred a very high scoring image over a paired low scoring image.

Our results are only partially consistent with those of Stoddard~\cite{stoddard2015popularity}. He concluded that vote counts in several subreddits were quite informative for items that got at least a few votes, whereas we find that they are only good at distinguishing the highest scoring from the lowest. Stoddard's method was to fit a Poisson model, then infer an underlying quality attribute of each item as a prediction of its votes. In the absence of ground truth values, however, it is not clear how good the model fit has to be in order to conclude that the social media vote counts are informative about user preferences.

\begin{figure}[t]
    \centering 
    
    \begin{tikzpicture}
\sffamily
\begin{axis}[ 
    title = {\small Reddit Vote-Scores},
    title style={align=center,yshift=-.1in}, 
    legend style={font=\scriptsize},
    width = 2.75in, height = 1.75in,  
    ylabel near ticks,
    xlabel near ticks,
    ylabel = {\small Reddit Score-Percentile $\Delta$}, 
    ymax=1.1, ymin=-.1,
    every tick label/.append style={font=\scriptsize},
    xmin=-0.1,xmax=1.05, 
    domain=-0.1:2,  
    xlabel={\small $\kappa$},
    xlabel style = {yshift=0.05in},
    ] 
     
    \addplot[only marks,mark options={scale=.55}, green, fill=green!30!white] coordinates { 
(-0.052631578947368474,0.07317927099999999)
(-0.04210526315789476,0.026744186047000018)
(-0.040000000000000036,0.023255813953000026)
(-0.0292397660818714,0.07297447200000007)
(-0.02857142857142858,0.08333333399999998)
(-0.02564102564102566,0.09999999999999998)
(-0.02564102564102566,0.03501561700000011)
(-0.02564102564102566,0.3164556960000001)
(-0.022727272727272707,0.03695652199999999)
(-0.020242914979757054,0.14505494499999994)
(-0.020242914979757054,0.742424242)
(-0.019607843137254943,0.022853747777999978)
(-0.018181818181818188,0.13991163475699997)
(-0.018181818181818188,0.009367828932000055)
(-0.016129032258064502,0.04761904800000005)
(-0.015819209039548032,0.0011261261260000488)
(-0.01548821548821544,0.03953370501799991)
(-0.01538461538461533,0.07633018398800007)
(-0.01538461538461533,0.468954249)
(-0.01538461538461533,0.012987013000000047)
(-0.015335801163405605,0.02285078487899983)
(-0.014207650273224015,0.10955966899000003)
(-0.014207650273224015,0.007061502164000033)
(-0.014207650273224015,nan)
(-0.013664596273291973,0.298323471401)
(-0.013664596273291973,0.006058739877000008)
(-0.013664596273291973,0.15189873417700006)
(-0.013461538461538414,0.03475402117899995)
(-0.012429378531073398,0.05016447368400001)
(-0.012162876784770016,0.01780626780600003)
(-0.011891891891891881,0.1041426927500001)
(-0.011180124223602483,0.06105577763000003)
(-0.010954616588419452,0.018181818182000042)
(-0.010736764161421664,0.34446627373899996)
(-0.010526315789473717,0.02164502164500004)
(-0.010526315789473717,0.17307692307699996)
(-0.010526315789473717,0.04545454545499994)
(-0.010196078431372602,0.00043815222499998807)
(-0.00983606557377048,0.09125116975299996)
(-0.00983606557377048,0.15429098314500012)
(-0.009615384615384581,0.008830022075000055)
(-0.009615384615384581,0.277086383602)
(-0.009446693657219951,0.09877253599999991)
(-0.009009009009009028,0.14511041009400005)
(-0.008403361344537785,0.559322034)
(-0.008403361344537785,0.050000000000000044)
(-0.007575757575757569,0.014285714000000005)
(-0.007039337474120111,0.012584704743999975)
(-0.006993006993006978,0.570238095238)
(-0.00694444444444442,0.10397946084800014)
(-0.00687466948704385,0.053571428570999946)
(-0.00687466948704385,0.032459016393000084)
(-0.00687466948704385,0.008417508418000041)
(-0.006779661016949157,0.03448275862100003)
(-0.004566210045662156,0.06578947368400001)
(-0.002020202020202033,0.026503169463999998)
(-0.002020202020202033,0.05959302325600002)
(-0.001851166234727919,0.046732995379)
(-0.001851166234727919,0.01891906341299998)
(-0.001851166234727919,0.36037966932)
(-0.0015649452269170805,0.05813953488400003)
(-0.001242236024844745,0.04876718448400008)
(-0.0004662004662004948,0.185499316006)
(0.0005288207297726721,0.03278688524499995)
(0.0005288207297726721,0.638888888889)
(0.0006835269993163973,0.17198905109499996)
(0.001129943502824915,0.24932713481800006)
(0.001129943502824915,0.159090909091)
(0.0022948938611588865,0.10701023081000005)
(0.0022948938611588865,0.07348798016800007)
(0.0038461538461538325,0.026280323449999976)
(0.0039525691699604515,0.050000000000000044)
(0.0039525691699604515,0.24675324675300014)
(0.0039525691699604515,0.0)
(0.0048130322102923895,0.0024968789020000104)
(0.005464480874316946,0.077351010316)
(0.006211180124223503,0.10465116279100006)
(0.006428988895382792,0.03333333333300004)
(0.006747638326585648,0.09615384599999999)
(0.008288288288288204,0.009628610729000009)
(0.010047593865679438,0.04678906716899989)
(0.0117647058823529,0.07692307700000012)
(0.0117647058823529,0.19999999999999996)
(0.014084507042253502,0.1037548001709998)
(0.015151515151515138,0.1593026030000001)
(0.015151515151515138,0.0)
(0.015320910973084967,0.17198905109499996)
(0.015320910973084967,0.030303030302999967)
(0.01538461538461533,0.04761904761900004)
(0.01754385964912286,0.02611832599999997)
(0.01754385964912286,0.040000000000000036)
(0.018118059614260718,0.04646660212999998)
(0.018181818181818077,0.08095378274999998)
(0.022222222222222143,0.016041162227000028)
(0.023693605972086917,0.20559246954600008)
(0.023728813559322104,0.03515132408499999)
(0.023728813559322104,0.01793032786899995)
(0.02499999999999991,0.012015177066000193)
(0.025290498974709585,0.03667702307099996)
(0.027450980392156765,0.011478368130000027)
(0.028340080971659853,0.061663652999999985)
(0.028919330289193246,0.04166666666700003)
(0.032258064516129004,0.8235294120000001)
(0.032258064516129004,0.08333333300000001)
(0.032738095238095344,0.005417118093000051)
(0.03407984420642651,0.10176565008100003)
(0.03442340791738374,0.05911623884999995)
(0.03543098889476459,0.08113425442699995)
(0.03599374021909241,0.014244186046999952)
(0.03838383838383841,0.080862533693)
(0.03850931677018643,0.025000000000000022)
(0.03850931677018643,0.028877005346999995)
(0.040000000000000036,0.16666666666700003)
(0.042105263157894646,0.038461538461999956)
(0.04428904428904423,0.019696969697000077)
(0.045324153757888785,0.13547340665500007)
(0.04627450980392167,0.013380135136000004)
(0.04627450980392167,0.05852667595399996)
(0.04761904761904767,0.013096351730999989)
(0.04850964348334297,0.004467683726000038)
(0.04850964348334297,0.30575964008799994)
(0.048530416951469535,0.07604348471299982)
(0.05129561078794298,0.01469756924799992)
(0.05138339920948609,0.04761904761900004)
(0.05138339920948609,0.040000000000000036)
(0.05138339920948609,0.06593406593399986)
(0.05153153153153145,0.19839773798299998)
(0.05153153153153145,0.10996749729200006)
(0.05258799171842643,0.043140485588999966)
(0.05258799171842643,0.023255813953000026)
(0.05258799171842643,0.036363636362999885)
(0.05384615384615388,0.13396778916599983)
(0.056189640035118504,0.1945440198120001)
(0.05723905723905731,0.01630206838699999)
(0.058461538461538565,0.09433962200000001)
(0.05882352941176472,0.014285714999999977)
(0.05882352941176472,0.06849314999999989)
(0.062200956937799035,0.11907066795699994)
(0.06832298136645965,0.09534671532899985)
(0.06832298136645965,0.22151898734199993)
(0.07115384615384612,0.09553072625699999)
(0.07115384615384612,0.02883875013199999)
(0.07457627118644061,0.032258064515999996)
(0.08333333333333326,0.310551558753)
(0.08333333333333326,0.036666666999999986)
(0.08333333333333326,0.1499999999999999)
(0.08333333333333326,0.037735849000000043)
(0.08333333333333326,0.157692308)
(0.08571428571428563,0.159015359656)
(0.08675799086757996,0.125)
(0.08771929824561409,0.04545454500000001)
(0.08937070333157049,0.031941031941000064)
(0.09333333333333327,0.03846153846100009)
(0.09333333333333327,0.261904761904)
(0.09604519774011289,0.028153153153000043)
(0.0990990990990992,0.015105164620999978)
(0.10476190476190483,0.01748598448399996)
(0.10502283105022836,0.016129032257999998)
(0.11158117398202005,0.07685810810799998)
(0.11158117398202005,0.013513513513999986)
(0.11163062536528345,0.21428571428600007)
(0.11260504201680677,0.19999999999999996)
(0.11260504201680677,0.633333333)
(0.11384615384615393,0.13235294200000003)
(0.11384615384615393,0.1428571428569999)
(0.11578947368421044,0.0)
(0.11888111888111896,0.12991944764099994)
(0.12786885245901636,0.03830581295400004)
(0.1287878787878789,0.35)
(0.1287878787878789,0.542372881)
(0.13590692755156009,0.10810810810799998)
(0.13734658094681462,0.025000000000000022)
(0.1384615384615384,0.25)
(0.1457627118644067,0.00620347394599996)
(0.1461187214611872,0.1641791044780001)
(0.14782608695652177,0.08644221059400004)
(0.15961538461538471,0.13103953147900005)
(0.15999999999999992,0.27272727272700004)
(0.1619718309859155,0.12467532467499987)
(0.16234796404019036,0.11545103709799998)
(0.16234796404019036,0.02238358985400002)
(0.16532258064516125,0.05714285699999999)
(0.1729323308270676,nan)
(0.18128654970760238,0.038461538461999956)
(0.18128654970760238,0.125)
(0.18153846153846165,0.04523809523799993)
(0.18205128205128207,0.043290042999999945)
(0.18205128205128207,0.664961637)
(0.1827027027027026,nan)
(0.19579193454120403,0.16363636363599987)
(0.19617224880382778,nan)
(0.2045197740112994,0.011549925484000023)
(0.20734126984126977,0.09279918864000014)
(0.20934761441090566,0.028329809724999988)
(0.21727395411605932,0.04761904800000005)
(0.22157588577472231,0.03448275862100003)
(0.22331002331002336,0.128842676311)
(0.22331002331002336,0.023255813953000026)
(0.2248447204968944,0.07142857142900005)
(0.226936026936027,0.024598914084000056)
(0.24,0.0)
(0.24242424242424243,0.08333333300000001)
(0.24657534246575352,0.09836065573800001)
(0.26054054054054054,0.26734348561800003)
(0.2615384615384615,0.040404039999999974)
(0.2615384615384615,0.038461538461999956)
(0.2615384615384615,0.23076923076900002)
(0.2635885447106956,0.0)
(0.27231638418079096,0.0)
(0.27231638418079096,0.879393355802)
(0.2846153846153847,0.153846154)
(0.28671328671328666,0.03842676310999982)
(0.28724440116845185,0.1511627906980001)
(0.28853754940711474,0.050000000000000044)
(0.305050505050505,0.3885585085919999)
(0.3055555555555556,0.06217882836599997)
(0.30980392156862746,0.10402405000000003)
(0.31060606060606055,0.09959046999999999)
(0.31580655631288534,0.012499999999999956)
(0.31677018633540377,0.13980022986999985)
(0.3262032085561497,0.08860759500000004)
(0.3262823902696985,0.04918032786899995)
(0.3262823902696985,0.21428571428499998)
(0.33063427800269896,0.07534246600000005)
(0.33333333333333326,0.06818181818100011)
(0.3407364114552893,nan)
(0.3491525423728814,0.11508771929900008)
(0.3538461538461539,0.550909090909)
(0.37225636523266026,0.10055464834700012)
(0.38636363636363646,0.14285714299999996)
(0.3954116059379218,0.11666666699999995)
(0.3954116059379218,0.0036231890000000266)
(0.3954116059379218,0.050000000000000044)
(0.40639269406392686,0.363636363637)
(0.4066631411951349,0.18032786885199992)
(0.4076923076923078,0.559322034)
(0.41512605042016815,0.08695652200000004)
(0.4385964912280702,0.012987013000000047)
(0.4385964912280702,0.032258064515999996)
(0.4490929285449834,0.12875121006800017)
(0.4500264410364887,0.06912108310800003)
(0.4500264410364887,0.06666666666700005)
(0.4556451612903225,0.166666667)
(0.4584615384615385,0.04545454500000001)
(0.4584615384615385,0.0)
(0.4584615384615385,0.11309523799999999)
(0.4696969696969697,0.12770562799999996)
(0.4813559322033898,0.062100213220000144)
(0.49550502379693273,0.22743589743600012)
(0.5411605937921729,0.04187437599999999)
(0.5411605937921729,0.196969697)
(0.5430988894764675,0.10267857142900005)
(0.6023391812865497,0.15384615384599998)
(0.6023391812865497,0.1428571428569999)
(0.6023391812865497,0.3333333333330001)
(0.6040404040404039,0.24007744433700007)
(0.6210526315789473,0.807692307692)
(0.622693096377307,0.011217736516999977)
(0.6590909090909092,0.752704577)
(0.6773109243697479,0.853050847)
(0.7046153846153846,0.125)
(0.7153846153846153,0.330952381)
(0.7153846153846153,0.030769230999999952)
(0.7468926553672317,0.34294436906400017)
(0.8461538461538463,0.13235294100000017)
(0.8823529411764706,0.181818182)
(1.0,0.691040924)
(1.0,0.4821801639999999)
    };
    
    \addplot[only marks, mark=square*, mark options={scale=.6}, red, opacity=.8, fill=red!30!white] coordinates { 
    (-0.04347826086956519,0.15476190476099994)
(-0.04210526315789476,0.015151515150999884)
(-0.03384615384615386,0.2142857142850001)
(-0.030303030303030276,0.2307692309999999)
(-0.0292397660818714,0.259097035)
(-0.0292397660818714,0.235294118)
(-0.0292397660818714,0.04761904761900004)
(-0.026666666666666616,0.050000000000000044)
(-0.02564102564102566,0.05869603000000001)
(-0.02564102564102566,0.009164420000000062)
(-0.023076923076923106,0.05476190400000003)
(-0.020242914979757054,0.21739130400000006)
(-0.018181818181818188,0.30258173042399994)
(-0.016470588235294126,0.1586186300150001)
(-0.016042780748663055,0.11688311699999998)
(-0.01548821548821544,0.18014022888300008)
(-0.01548821548821544,0.01634382566600001)
(-0.01538461538461533,0.032258064515999996)
(-0.015335801163405605,0.0016175342330000397)
(-0.015335801163405605,0.02473546744099997)
(-0.014047410008779626,0.005743395401999973)
(-0.014047410008779626,0.527279273291)
(-0.012429378531073398,0.1372377622369999)
(-0.011413520632133412,0.16307106859100007)
(-0.011413520632133412,0.06827242524900001)
(-0.011180124223602483,0.22561271717199993)
(-0.010954616588419452,0.02502207830400005)
(-0.009935710111046214,0.01163639789899995)
(-0.009935710111046214,0.2538427221230001)
(-0.009935710111046214,0.06666666666599996)
(-0.009935710111046214,0.023255813952999915)
(-0.00983606557377048,0.12707762561699998)
(-0.009753298909925379,0.0025265037930000123)
(-0.009446693657219951,0.02857142899999998)
(-0.0091324200913242,0.02491554054099998)
(-0.0070237050043898686,0.05356206784600004)
(-0.00687466948704385,0.0625)
(-0.006779661016949157,0.02141203703700001)
(-0.003278688524590123,0.02578715924399999)
(-0.002020202020202033,0.02228810879200005)
(-0.002020202020202033,0.129498226052)
(-0.0015649452269170805,0.030303030302999967)
(-0.0007843137254901489,0.013685695503000028)
(0.0,0.02346941086300003)
(0.0017825311942958333,0.0)
(0.0017825311942958333,0.00988319900000001)
(0.0040322580645162365,0.050000000000000044)
(0.0048130322102923895,0.033684210526000014)
(0.006211180124223503,0.03550512445099996)
(0.006428988895382792,0.0690643185369999)
(0.006428988895382792,0.09302325581399995)
(0.006815968841285214,0.004761904762000002)
(0.0070237050043897575,0.02143423103099995)
(0.015151515151515138,0.07534246600000005)
(0.016064257028112428,0.0051354577579999505)
(0.018118059614260718,0.028571428571000035)
(0.018118059614260718,0.01424300174899995)
(0.0182648401826484,0.038461538461999956)
(0.01984126984126977,0.2260778128280001)
(0.023728813559322104,0.07388663967600007)
(0.025290498974709585,0.34326019358899995)
(0.027027027027026973,0.029610829102999947)
(0.027450980392156765,0.005353430835999928)
(0.028094820017559252,0.037068183304999947)
(0.030303030303030276,0.11392405063300015)
(0.03214494447691418,0.0237082873240001)
(0.032258064516129004,0.0)
(0.03442340791738374,0.24559032976700002)
(0.03442340791738374,0.044628042871000084)
(0.03599374021909241,0.024472573839000034)
(0.03855855855855861,0.09275184275199999)
(0.041095890410958846,0.11363636363599994)
(0.0412642669007901,0.13121316643100012)
(0.042105263157894646,0.0)
(0.042105263157894646,0.13636363636399995)
(0.04428904428904423,0.018181818182000042)
(0.04480874316939887,0.0137609767789999)
(0.045324153757888785,0.03296801571399999)
(0.04761904761904767,0.018540137464000073)
(0.05129561078794298,0.17857142857099995)
(0.05258799171842643,nan)
(0.05258799171842643,0.14844941535299994)
(0.06034801925212885,0.04715672676900007)
(0.06416275430359941,0.08896103896100005)
(0.0730282375851996,0.007751937983999979)
(0.07878787878787885,0.10096025919499996)
(0.08571428571428563,0.17107269954300008)
(0.08860759493670889,0.03749999999999998)
(0.09333333333333327,0.0)
(0.09859154929577474,0.013095238095000017)
(0.10040160642570273,0.024299457767000043)
(0.11163062536528345,0.0054590856709999525)
(0.11163062536528345,0.23595505617900003)
(0.11260504201680677,0.02857142899999998)
(0.11384615384615393,0.2705244130000001)
(0.11462450592885376,0.02097902097899995)
(0.1287878787878789,0.08333333300000001)
(0.1384615384615384,0.184782609)
(0.14782608695652177,0.002697760943000005)
(0.15294117647058814,0.13062318424600006)
(0.1595959595959595,0.004617304606000072)
(0.16532258064516125,0.0769230769999999)
(0.17184265010351973,0.01685393258500001)
(0.1729323308270676,0.013108614232000049)
(0.18128654970760238,0.040000000000000036)
(0.18181818181818188,0.156924795)
(0.1909042834479111,0.04426213412400004)
(0.20029618659755655,0.01244269810099996)
(0.20934761441090566,0.05151515151500008)
(0.22077922077922074,0.2745853927080001)
(0.226936026936027,0.0221511183580001)
(0.25383022774327113,0.06329113924100005)
(0.28248796741947424,0.042567125082)
(0.28724440116845185,0.32675725111)
(0.29824561403508776,0.0)
(0.29824561403508776,0.050000000000000044)
(0.305050505050505,0.10351089588399986)
(0.3096045197740114,0.042207792206999994)
(0.31060606060606055,0.02857142899999998)
(0.3262032085561497,0.33544303799999997)
(0.33333333333333326,0.0)
(0.5256916996047432,0.019230769230999978)
(0.6684491978609626,0.11392405100000014)
(0.7046153846153846,0.03692307692300001)
(0.7894736842105263,0.10714285714300009)
(1.0,0.033613444999999964)
    };

\end{axis}    
\end{tikzpicture}  


    \begin{tikzpicture} 
    \begin{axis}[%
    hide axis, scale only axis,
	height=.25in, width=2in,
    xmin=0,xmax=50,ymin=0,ymax=0.4,
    legend style={at={(1.5,0)},draw=none,fill=none, legend cell align=left,legend columns = -1, column sep = 1mm}
    ]
    \addlegendimage{only marks,mark options={scale=1}, green, fill=green!30!white}
    \addlegendentry{\scriptsize Correct Prediction};
    \addlegendimage{mark=square*,only marks,mark options={scale=1}, red, fill=red!30!white}
    \addlegendentry{\scriptsize Incorrect Prediction};
    \end{axis}
    \end{tikzpicture}

\vspace{-0.1in}

    \caption{Reddit score-percentile $\Delta$ of image-pairs as a function of   groundtruth agreement ($\kappa$). Pairs for which Reddit vote-scores correctly predicted the majority opinion are denoted with green $\circ$ with incorrect predictions represented with red $\square$.}
    \label{fig:type_of_error}
\end{figure}
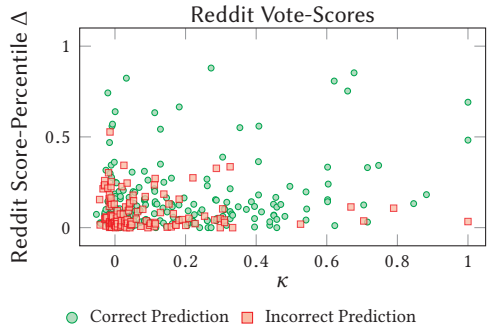

Our results also differ somewhat from those of Figueiredo et al~\cite{Figueiredo:2014:CDI:2611105.2557285}. Their findings for YouTube suggested that a huge difference in views was not a reliable indicator of consensus preference for one video over another. By contrast, we find that a huge difference in Reddit score or Imgur views is predictive of the majority's true preference (Fig.~\ref{fig:pref_accuracy_pairlogic}, Fig.~\ref{fig:pref_accuracy_scorediff_corr} and Fig.~\ref{fig:type_of_error}). It's not clear why the results differ. While our game attempts to make voluntary participation more attractive, the structure of the task itself was similar between our studies: subjects offered an opinion about which of a pair of items they preferred. One possible explanation of the different results is that there are algorithmic and design differences between YouTube and Reddit, with extreme differences in popularity of scores for images on Reddit being more informative of popular opinion. Another possibility is differences in the items being evaluated: there may be taste variation for the YouTube videos assessed, such that topical preferences outweighed any assessment of production quality. By contrast, even though some people may prefer cat images to dog images, it may be that most people appreciate any really high quality image.
This is a clear avenue for additional research. Additionally, Figueiredo et al also found that, contingent on one video being strongly preferred over another, the preferred one almost always had more views. Our results show a similar finding for $\kappa\ge$ .40, with the added complication that only 10\% of the image-pairs had an agreement larger than 0.40. Even then, near perfect agreement still resulted in incorrect predictions some of the time. These findings are illustrated in Figure~\ref{fig:type_of_error}, which plots the score-percentile $\Delta$ as a function of the agreement level with correct and incorrect predictions in green {\Large $\circ$} and red $\square$, respectively.

The Reddit scores, it seems, provide a very useful signal for the casual users who only look at the top few images in a subreddit. They are reliably better (i.e., preferred by more people) than low scoring images. Differences within the top quartile of scores, however, are not meaningful.
This suggests two potential design implications. First, if score differences between adjacent posts are not meaningful, it might be better not to show them at all. Second, it might be valuable to partially randomize the order so as to dampen social influence effects. 

We found that image-pairs sourced from smaller subreddits were better predictors of GuessTheKarma majority vote than image-pairs sourced from larger subreddits. One plausible explanation is that, because of their size and the large volume of posts, large subreddits are more prone to algorithmic and social biases. Larger subreddits rely heavily on Reddit's algorithm to surface the most informative or interesting posts. This may create a ranking bias effect. A post that garnered few votes and little attention within a large subreddit will be crowded out by more popular posts and is not likely to be seen by many users. However, in a small subreddit with little volume or competition, ranking bias will have a far smaller effect. In addition, larger subreddits' posts are more frequently included on Reddit's frontpage, which may create a strong herding effect. Further study is needed to thoroughly understand the relationship between the size of an online community and its ability to solicit user preferences.

One should interpret the findings and design recommendations with some caution. First, while we had many pairs of images with similar Reddit scores, we had few pairs with very different Reddit scores, the VH-L and H-L combinations. In a larger sample of image pairs, the correlation of Reddit score difference with true population preference might not hold up. 
Second, our ground truth proxy from the GuessTheKarma players may not always reflect the true preference of the majority of readers of the particular subreddits. We tried to recruit Redditors, but many of them came from a games subreddit. We deliberately chose image subreddits where specialized knowledge was not important, but our players' tastes may not have matched readers of the image subreddits. In addition, game players judged images months, sometimes years, after they were originally posted on Reddit. Tastes could have changed in the interim; this seems unlikely for dog and cat pictures, but might be the case in, say, /r/CrappyDesign.
Third, it is possible that a user disregards the post image and title and instead perceives quality from the social engagement of the comment section. However, recent reports suggest that a post is rarely up/down-voted after reading the comments section~\cite{glenski2017consumers}.

A critical feature of GuessTheKarma as a method of collecting data is that the game had to be fun in order to attract voluntary participation. We think two things made this game fun. First, the challenge level was just right, neither too easy nor too hard. Indeed, in pilot tests people complained about some pairs that were too hard to judge, and these turned out to be pairs where both were in the bottom half of Reddit scores. Once we removed L-L comparisons, we got fewer complaints. Second, the images were generally interesting or entertaining to look at. A nice side effect of eliminating the L-L comparisons was that at least one of every pair had received many upvotes and thus was probably not terrible. Thus, participating in the game was in part a discovery activity comparable to browsing one of the image subreddits.

This, however, suggests an important limitation of GuessTheKarma as a data collector. It may work only for pairs of items that are a) interesting to interact with in their own right, and b) somewhat but not too challenging to guess the popular score for. If the information signal from a social popularity metric is too low, then trying to guess the metric will be too challenging, and people won't participate. If the information signal from a social popularity metric is too high, then trying to guess the metric will be too easy, and people won't participate. 
Thus, our finding that the Reddit scores provide only coarse-grained information but not fine-grained information about true popularity may be the only state of the world where we could have gathered data using GuessTheKarma. If, in fact, Reddit scores provided fine-grained information or no information at all, we might not have been able to attract participants to play the game voluntarily. 

It is important to note that the predictions collected in the experiment described above are binary predictions; participants either chose image A or image B as their preference. As a result, a player's confidence in their choice is not indicated in their judgment. Future work on this task should explore alternative designs that allow more information to be collected from players. These designs would allow players to include a degree of confidence in their choices, to estimate the difference between image scores, or to ``pass'' on image-pairs that they are not confident in. Future designs might also be adapted to solicit a measure of quality to allow direct comparisons between quality and popularity. Finally, it may be possible to allow players to win small cash prizes and even wager their earnings to increase their reward.

\section{Conclusions}

In this work we presented GuessTheKarma, a survey tool that can be used to explore the performance of social rating systems, and an analysis of two social media rating systems, Reddit votes and Imgur views. Through its tournament style setup, the GuessTheKarma game removed many of the social influence effects and algorithmic, design, and ranking biases, that have been previously reported in social rating systems research. Therefore, discrepancies between GuessTheKarma-judgments and Reddit scores (or Imgur view counts) can be primarily explained by the presence of socio-technical influence dynamics in the the Reddit scores and Imgur view counts.

The results suggest that Reddit vote-scores and Imgur view-counts are relatively poor predictors of user preference except when one image has a much higher score than the other. For example, images with scores in the top 5\% of scores within their subreddit paired with an image whose score falls within the bottom half of all scores were preferred by GuessTheKarma players almost 90\% of the time.
Unsurprisingly, we found that individual players were less accurate predictors, performing slightly better than random. However, we also did not find significant differences between players of varying expertise or familiarity with the platforms or communities we compared. Finally, we found no difference in the accuracy of active members of a subreddit, even small subreddits, compared to those respondents that did not use Reddit.

\section*{Acknowledgements}
The authors gratefully acknowledge support by the Defense Advanced Research Projects Agency (DARPA \#W911NF-17-C-0094). The views and conclusions contained herein are those of the authors and should not be interpreted as necessarily representing the official policies or endorsements, either expressed or implied, of DARPA or the U.S. Government.

\bibliographystyle{ACM-Reference-Format}
\bibliography{gtk}

\end{document}